\documentclass[aps,twocolumn,showpacs,preprintnumbers]{revtex4}

\usepackage[utf8]{inputenc}
\usepackage[english]{babel}
\usepackage{amsmath}
\usepackage{braket}
\usepackage{bm}
\usepackage{chemformula}
\usepackage{siunitx}
\usepackage{amsbsy}

\usepackage{graphicx}
\usepackage{amssymb}
\usepackage{amsfonts,dsfont}
\usepackage{subfigure}

\usepackage[normalem]{ulem}
\usepackage[bbgreekl]{mathbbol}
\usepackage{mathrsfs,empheq}
\usepackage{relsize,scalerel}

\newcommand{\pstmmc}{$\textit{Pm}\overline{3}\textit{n }$}
\newcommand{\AlH}{AlH\textsubscript{3} }

\newcommand{\Rcal}{\mathcal{R}}

\newcommand{\Fcal}{\mathcal{F}}

\newcommand{\bR}{\boldsymbol{R}}

\newcommand{\bD}{\boldsymbol{D}}

\newcommand{\bG}{\boldsymbol{G}}

\newcommand{\bq}{\boldsymbol{q}}
\newcommand{\bk}{\boldsymbol{k}}

\newcommand{\bvarPhi}{\boldsymbol{\varPhi}}

\newcommand{\bRcal}{\boldsymbol{\Rcal}}

\newcommand{\bbOmega}{\mathbb{\Omega}}

\newcommand{\sss}[1]{\scriptscriptstyle{\text{#1}}}

\newcommand{\rscha}{\bRcal}
\newcommand{\rschatrial}{\bRcal}
\newcommand{\phischatrial}{\bvarPhi}

\newcommand{\rhoschatrial}{{\tilde\rho}_{\scriptscriptstyle{\rscha},\scriptscriptstyle{\phischatrial}}}

\newcommand{\Rcaleq}{\Rcal}
\newcommand{\bRcaleq}{\bRcal}
\newcommand{\DF}{D^{\sss{(F)}}}
\newcommand{\bDF}{\bD^{\sss{(F)}}}

\renewcommand{\Im}{\mathrm{Im}}

\begin{document}

\title{Strong Anharmonic and Quantum Effects in \pstmmc  \AlH  Under High Pressure:\newline A First-Principles Study}

\author{Pugeng Hou$^{1}$, Francesco Belli$^{2,3}$, Raffaello Bianco$^{3}$, Ion Errea$^{2,3,4}$}

\affiliation {$^1$College of Science, Northeast Electric Power University, Changchun Road 169, 132012, Jilin, P. R. China}

\affiliation {$^2$Fisika Aplikatua 1 Saila, Gipuzkoako Ingeniaritza Eskola, University of the Basque Country (UPV/EHU), Europa Plaza 1, 20018 Donostia/San Sebasti\'an, Spain}

\affiliation {$^3$Centro de F\'isica de Materiales (CSIC-UPV/EHU), Manuel de Lardizabal Pasealekua 5, 20018 Donostia/San Sebasti\'an, Spain}

\affiliation {$^4$Donostia International Physics Center (DIPC), Manuel de Lardizabal Pasealekua 4, 20018 Donostia/San Sebasti\'an, Spain}

\date{\today}

\begin{abstract} 
Motivated by the absence of experimental superconductivity in 
the metallic \pstmmc phase of \AlH despite the predictions, 
we reanalyze its vibrational and supeconducting properties at pressures $P \ge 99$ GPa making use of first-principles techniques. In our calculations based on the
self-consistent harmonic approximation method that treats anharmonicity beyond
perturbation theory, we predict a strong anharmonic correction to the phonon spectra and demonstrate that the superconducting critical temperatures predicted in previous calculations based on the harmonic approximation are strongly suppressed by anharmonicity.
The electron-phonon coupling concentrates on the lowest-energy hydrogen-character optical modes at the X point of the Brillouin zone. As a consequence of the strong anharmonic enhancement of their frequency, the electron-phonon coupling is suppressed by at least a 30\%.
The suppression in $\lambda$ makes $T_c$ smaller than 4.2 K above 120 GPa, which is well consistent with the experimental evidence. Our results underline that metal hydrides with hydrogen atoms in interstitial sites are subject to huge anharmonic effects.
\end{abstract}

\maketitle

\section{Introduction}

Motivated by the quest for metallic and superconducting hydrogen at very high pressures\cite{1}, a combination of first-principles structural predictions and calculations of the electron-phonon interaction has led in the last years to the prediction of many superconducting hydrides with high values of the superconducting critical temperature
($T_c$)\cite{FLORESLIVAS20201,BI2019,doi:10.1146/annurev-conmatphys-031218-013413,2,3,4,5,6,7,8,9,10,11,12,13,14,15,16,17,18}. 
Even if the observation of high-$T_c$ in pure hydrogen remains still elusive, although optical evidences of the probably superconducting atomic phase\cite{H:Borinaga_PRB_2016,borinaga2018strong} have been reported\cite{Dias_hydrogen_Science2017}, it is now an experimental fact that room temperature superconductivity is possible in hydrogen-rich ``superhydride'' compounds. Critical temperatures above 200 K have been observed in sulfur\cite{DrozdovEremets_Nature2015}, lanthanum\cite{Hemley-LaH10_PRL_2019,Nature_LaH_Eremets_2019}, and yttrium\cite{troyan2020anomalous,snider2020superconductivity,kong2019superconductivity} superhydrides at pressures exceeding 100 GPa. A mixture of C-H-S has finally reached room temperature superconductivity at pressures above 250 GPa\cite{Snider2020}, showing that there is lots of room for further increase of $T_c$ among ternary compounds\cite{16}. The role of theoretical first-principles calculations in all these experimental discoveries should be highlighted. For instance the discoveries of high-$T_c$ superconductivity in sulfur, 
lanthanum, and yttrium hydrides had been anticipated by {\it ab initio} calculations\cite{11,12,19,14}.


The standard procedure in these {\it ab initio} calculations relies on a classical treatment of the ions: the predicted structures are minima of the Born-Oppenheimer energy surface (BOES) and the phonons entering the superconducting equations are estimated assuming a harmonic expansion of the BOES around these crystal configurations. However, this classical (or harmonic) approach often completely breaks down as it neglects the quantum contribution from the kinetic term of the nuclei Hamiltonian to the energy and the phonon frequencies. The latter is large in hydrogen-based compounds due to the lightness of H atoms.
Consequently, the $T_c$ from classical harmonic calculations\cite{8,15,19} usually differ from the experimental values\cite{19,DrozdovEremets_Nature2015,Nature_LaH_Eremets_2019}. 
In fact, the anharmonic correction to the phonon frequencies imposed by the large ionic quantum fluctuations strongly renormalizes the superconducting critical temperatures in hydrogen-based superconductors, yielding $T_c$'s in close agreement with experiments\cite{22,23,24,25,26}. 
Furthermore, quantum anharmonic effects also explain the stabilization of the  
crystal structures of superhydrides observed experimentally, as, otherwise, these structures would not be the ground state\cite{24,25}.

In the literature of superconducting hydrides \AlH\ deserves a remarkable position as it was one of the first metallic hydrogen-based compounds synthesized at high pressures\cite{19} after been predicted theoretically by crystal structure prediction methods\cite{27}. Despite been predicted to be a superconductor at 24 K at 110 GPa in the \pstmmc phase within standard harmonic calculations, experimentally no superconductivity was observed down to 4 K over the 120–164 GPa pressure range\cite{19}. 
It was later suggested that anharmonicity was responsible for the suppression of $T_c$\cite{28}. 
Even if this seemed to close the debate on the experimental and theoretical disagreement, the perturbative treatment of anharmonicity followed in Ref. \cite{28} for this system seems questionable, as the anharmonic self-energy for some particular modes was estimated to be as high as the phonon frequencies themselves.
In these conditions perturbative approaches may lead to strong errors in the estimation of the renormalized phonon frequencies\cite{Errea2016}. Furthermore, anharmonic corrections were only estimated for few modes at only one pressure. A deeper analysis based not on a perturbative method is thus required to confirm that anharmonicity is responsible for the suppression of $T_c$ in \AlH.

In this work we present a thorough first-principles analysis of the full anharmonic phonon spectra of \pstmmc \AlH in a wide pressure range based on the variational 
stochastic self-consistent harmonic approximation (SSCHA) method\cite{22,29,30}. The calculated superconducting critical temperature is strongly suppressed by anharmonicity in the whole pressure range, in agreement with the absence of superconductivity in the 120-164 GPa pressure range below 4 K, confirming the suggestion made in Ref. \cite{28}. The paper is organized as follows: Sec. \ref{sec:methodolgy} describes the theoretical framework of our anharmonic {\it ab initio} calculations, Sec. \ref{sec:computational_details} overviews the computational details of our calculations, Sec. \ref{sec:results} presents the results of the calculations, and Sec. \ref{sec:conclusions} summarizes the main conclusions of this work.

\section{Methodology}
\label{sec:methodolgy}

\begin{figure}[t]
\includegraphics[width=1.0\columnwidth]{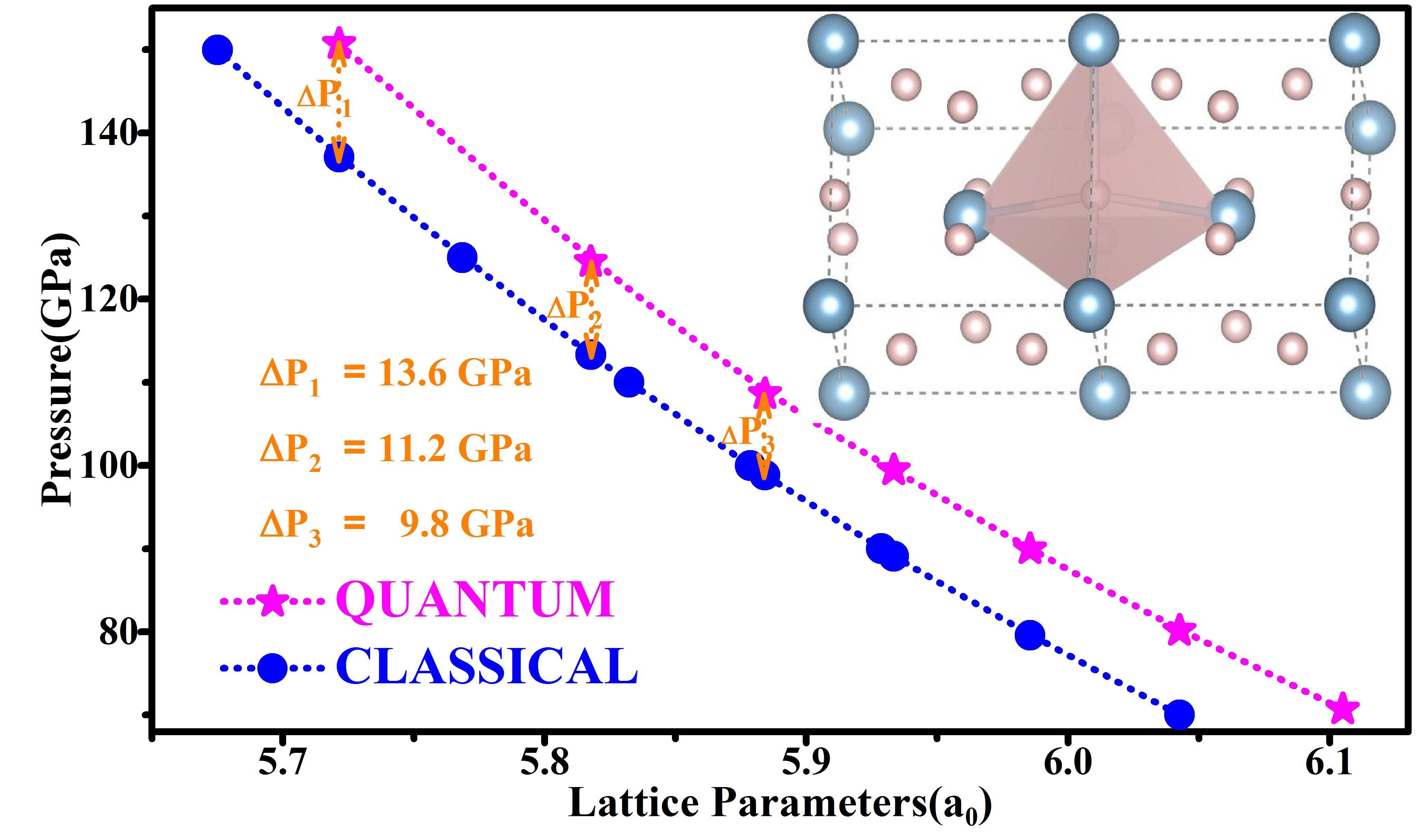}
\caption{(Color online) Comparison between the classical and quantum pressures as a function of the lattice parameter. The classical pressure is obtained from the BOES and the quantum from the SSCHA free energy. Here are shown the differences at three pressures for the same lattice parameter. The crystal structure of \pstmmc \AlH is shown as an illustration at the upper right corner, the blue spheres represent Al atoms and the pink spheres represent H atoms, respectively. One of the tetrahedra
surrounding hydrogen atoms is depicted. }
\label{1.png} 
\end{figure}

In the following, we briefly review the SSCHA method used for the calculation of anharmonic phonon frequencies, as well as the theoretical framework followed for estimating the superconducting critical temperature.

\subsection{The stochastic self-consistent harmonic approximation}

The SSCHA\cite{22,29,30} is a quantum variational method that minimizes the 
free energy of the system calculated with a trial density matrix $\rhoschatrial$:
\begin{equation}
    \Fcal[\rhoschatrial] = \braket{K + V(\bR)}_{\rhoschatrial} - TS[\rhoschatrial].
    \label{eq:sscha_f}
\end{equation}
Here, $K$ is the ionic kinetic energy, $V(\bR)$ the full Born-Oppenheimer potential,
$T$ the temperature, and $S[\rhoschatrial]$ the entropy calculated with the trial
density matrix. In the SSCHA the density matrix is parametrized with
centroid positions $\rschatrial$, which determine the average ionic positions, 
and  auxiliar force constants $\phischatrial$, which are related to the broadening
of the ionic wave functions around $\rschatrial$. Thus, minimizing $\Fcal[\rhoschatrial]$ with respect
to $\rschatrial$ and $\phischatrial$ a good variational approximation of the free energy
can be obtained without approximating the Born-Oppenheimer potential. 
This free energy can be used to estimate thermodynamic 
magnitudes, such as the pressure, including the effects of ionic quantum fluctuations\cite{30}.  
These effects are neglected if the pressure is estimated instead from $V(\bR)$, which
is the standard procedure. 

Phonon frequencies within the SSCHA should be calculated from the dynamical
extension of the theory\cite{29,monacelli2020time,lihm2020gaussian}. 
In this framework, phonon frequencies at the $\bq$ point of the
Brillouin zone (BZ) appear as peaks of the one-phonon 
spectral function
\begin{equation}
\sigma(\bq,\Omega) = -\frac{\Omega}{\pi}\,\Im\mathrm{Tr}\left[\bG(\bq,\Omega + i0^+)\right],
\label{eq:sigma}
\end{equation}
where $\bG(\bq,z)$ is the Fourier transform of the
Green's function for the variable $\sqrt{M_a}(R^a-\Rcaleq^a)$, which 
is related to the correlation between displacements
of atoms from the centroid positions. The index $a$ labels both an atom
and a Cartesian direction, and $M_a$ is the mass of atom $a$. $0^+$ is a small
positive number. We calculate the spectral
function both keeping the full energy dependence of the phonon self-energy and
within the so-called Lorentzian approximation (see Ref. \cite{PhysRevB.97.214101}
for details). In the latter case, the spectral function has well-defined Lorentzian
lineshape, with well-defined peaks at the $\bbOmega_{\mu}(\bq)$ frequencies.

In the $\Omega\to 0$ static limit, the peaks coincide with the $\Omega_\mu(\bq)$ frequencies, 
with $\Omega^2_\mu(\bq)$ being the eigenvalues of the Fourier 
transform of the free energy Hessian matrix
\begin{equation}
\DF_{ab}= \frac{1}{\sqrt{M_aM_b}} 
\left[ \frac{\partial^2F}{\partial\Rcal^a\partial\Rcal^b}\right]_{\bRcaleq}.
\label{eq:df}
\end{equation}
In Eq. \eqref{eq:df} $F$ is assumed to be the free energy at the minimum and $\bRcaleq$ the centroid
positions that minimize Eq. \eqref{eq:sscha_f}. As the phonon frequencies obtained
in this static limit are determined by the free energy Hessian, $\Omega_\mu(\bq)$ imaginary 
frequencies point to lattice instabilities in the quantum anharmonic energy landscape. 


\begin{figure*}[t]
\includegraphics[width=1.0\textwidth]{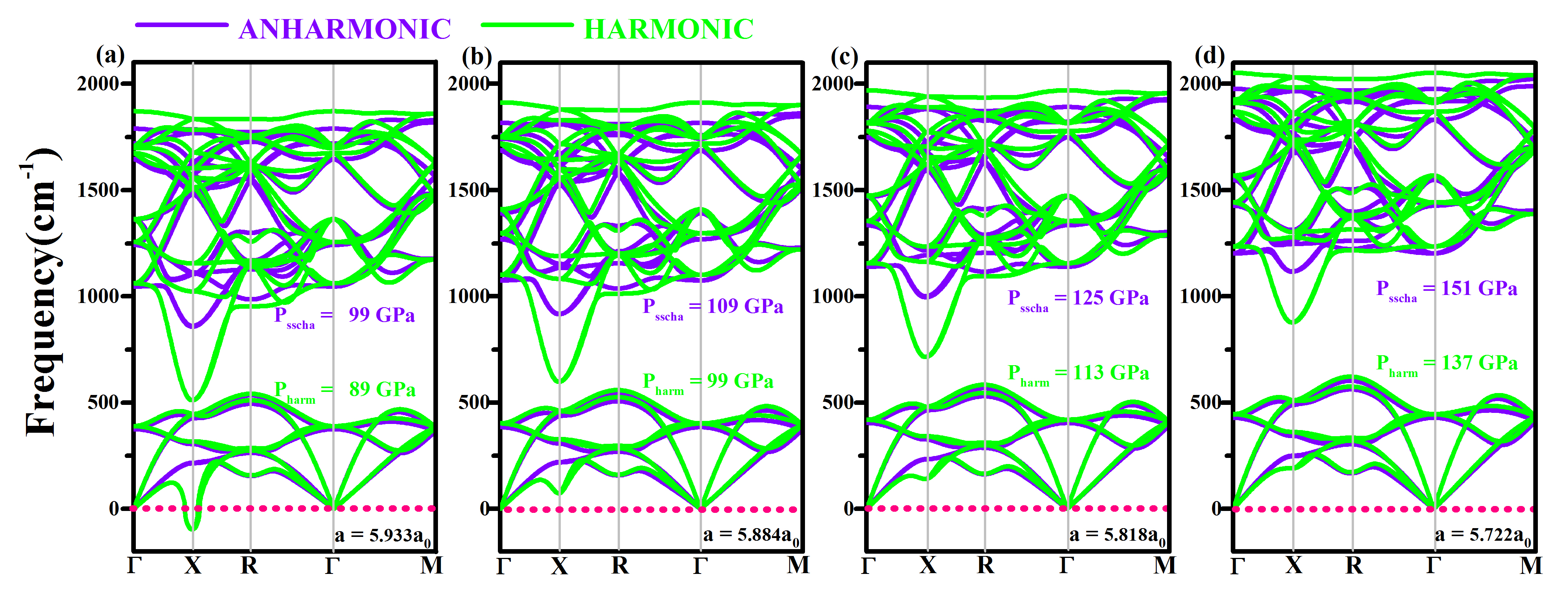}
\caption{(Color online) Comparison between the harmonic (green solid lines) and anharmonic (violet solid lines) phonon spectra of the cubic high-symmetry \pstmmc phase of \AlH for different lattice parameters. The anharmonic spectra are obtained from $\bDF$ and correspond to the static limit of the SSCHA dynamical theory.
The pressure calculated classically (harmonic calculation) 
and with quantum effects (anharmonic calculation) is marked in each case. The region of positive and negative frequencies, which represent imaginary frequencies, are separated with a pink dotted line.} 
\label{2.jpg} 
\end{figure*}

\subsection{Calculation of the superconducting transition temperature}

We evaluate $T_c$ with the Allen–Dynes\cite{31} modified McMillan equation,
\begin{equation}
T_c = \frac{f\textsubscript{1}f\textsubscript{2}\,\omega\textsubscript{log}}{1.2} \exp \left[ -\frac{1.04(1+\lambda)}{\lambda-\mu^*(1+0.62\lambda)} \right],
\label{eq1}
\end{equation}
where $\lambda$ is the electron-phonon coupling constant and $\mu^*$ is 
a parameter usually named as the Coulomb pseudopotential\cite{32}. This equation has
led $T_c$ values in rather good agreement with experiments in superhydrides\cite{25} despite its
simplicity.
$\lambda$ is defined as the first reciprocal moment of the electron-phonon Eliashberg function $\alpha^{2}F(\omega)$,
\begin{equation}
\lambda = 2 {\int_0^\infty d\omega \frac{\alpha^{2}F(\omega)}{\omega}}.
\label{eq2}
\end{equation}
Similarly
\begin{eqnarray}
    \omega\textsubscript{log} & = & \exp \left( \frac{2}{\lambda} \int d\omega \frac{\alpha^2F(\omega)}{\omega} \log\omega \right), \\
    f_1 & = & \left[ 1 + (\lambda / \Lambda_1)^{3/2} \right]^{1/3}, \\
    f_2 & = & 1 + \frac{(\bar{\omega}_2/\omega\textsubscript{log} - 1) \lambda^2}{\lambda^2 + \Lambda_2^2}
\end{eqnarray}
are also determined with $\alpha^{2}F(\omega)$. The $\Lambda_1$, $\Lambda_2$, and $\bar{\omega}_2$
parameters entering the equations above are given by
\begin{eqnarray}
    \Lambda_1 & = & 2.46 (1 + 3.8\mu^*) \\
    \Lambda_2 & = & 1.82 (1 + 6.3\mu^*)(\bar{\omega}_2/\omega\textsubscript{log}) \\
    \bar{\omega}_2 & = & \left[ \frac{2}{\lambda} \int d\omega \alpha^2F(\omega) \omega \right]^{1/2}.
\end{eqnarray}

We calculate the Eliashberg function as
\begin{equation}
    \alpha^{2}F(\omega) = \frac{1}{2\pi N(0) N_q} \sum_{\mu \bq} \frac{\gamma_{\mu}(\bq)}{\omega_{\mu}(\bq)} \delta(\omega -  \omega_{\mu} (\bq)),
\label{eq:eliashberg}
\end{equation}
where 
\begin{eqnarray}
    \gamma_{\mu}(\bq) & = &  \frac{\pi}{N_k} \sum_{{\bk}nm} \sum_{\bar{a}\bar{b}} 
      \frac{\epsilon_{\mu}^{\bar{a}}(\bq) \epsilon_{\mu}^{\bar{b}}(\bq)^*}{ \sqrt{M_{\bar{a}}M_{\bar{b}}}}
      d^{\bar{a}}_{{\bk}n,{\bk}+{\bq}m} d^{\bar{b}*}_{{\bk}n,{\bk}+{\bq}m} \nonumber \\ &\times&
      \delta(\varepsilon_{{\bk}n})
   \delta(\varepsilon_{{\bk+\bq}m})
   \label{eq:elphgamma}
\end{eqnarray}
is the phonon linewidth associated to the electron-phonon interaction of the mode
$\mu$ at wavevector $\bq$.
In Eqs. \eqref{eq:eliashberg} and \eqref{eq:elphgamma} $d^{\bar{a}}_{{\bk}n,{\bk}+{\bq}m} = \bra{{\bk}n}
\delta V_{KS} / \delta R^{\bar{a}}(\bq) \ket{{\bk}+{\bq}m}$, where 
$\ket{{\bk}n}$ is a Kohn-Sham state with energy $\varepsilon_{{\bk}n}$
measured from the Fermi level, $V_{KS}$ is Kohn-Sham potential, and
$R^{\bar{a}}(\bq)$ is the Fourier transformed displacement of atom
$\bar{a}$; $N_k$ and $N_q$ are the number of electron
and phonon momentum points used for the BZ sampling; $N(0)$ is the density of states 
at the Fermi level; and  $\omega_{\mu} (\bq)$ and 
$\epsilon_{\mu}^{\bar{a}}(\bq)$ represent phonon frequencies and polarization vectors.
The combined atom and Cartesian indexes with a bar ($\bar{a}$) 
only run for atoms inside the unit cell.
In this work, the Eliashberg function is calculated both at the harmonic or anharmonic levels,
respectively, by plugging into Eqs. \eqref{eq:eliashberg} and \eqref{eq:elphgamma} the harmonic phonon frequencies and polarization vectors or their anharmonic counterparts obtained diagonalizing $\bDF$.










\section{Computational Details}
\label{sec:computational_details}

Electronic properties are computed using density functional theory (DFT) 
as implemented in the \textsc{Quantum ESPRESSO}  package\cite{34,Giannozzi_2017}. 
Ultrasoft pseudopotentials\cite{35},
including $3$ electrons in the valence for 
Al, and a generalized gradient approximation 
for the exchange correlation potential are used\cite{36}. 
The plane-wave basis cutoff is set to 80 Ry and to 800 Ry for the density. 
Fist BZ integrations are performed on a 24$\times$24$\times$24 Monkhorst-Pack mesh, 
using a smearing parameter of 0.02 Ry. Harmonic phonon frequencies and electron-phonon matrix
elements entering Eq. \eqref{eq:elphgamma} are calculated within
density functional perturbation theory (DFPT)\cite{DFPT_S.Baroni}. 

The SSCHA variational minimization requires the calculation of forces in supercells. 
We calculate them within DFT in a 2$\times$2$\times$2 supercell containing 64 atoms, 
yielding dynamical matrices on a commensurate 2$\times$2$\times$2 grid. 
The difference between the harmonic and anharmonic dynamical matrices 
in the 2$\times$2$\times$2 grid was interpolated to a 13$\times$13$\times$13 grid. 
Adding the harmonic dynamical matrices in this fine grid to the result, 
the anharmonic dynamical matrices in the 13$\times$13$\times$13 grid are obtained.
Converging the value of the electron-phonon coupling constant required, indeed, 
a 13$\times$13$\times$13 $\bq$-point grid. 
A 60$\times$60$\times$60 $\bk$-point grid is used instead for the 
electronic integration in Eq. \eqref{eq:elphgamma} and the Dirac deltas are
approximated with Gaussian functions of 0.008 Ry width. 

\begin{figure}[t]
\includegraphics[width=1.0\columnwidth]{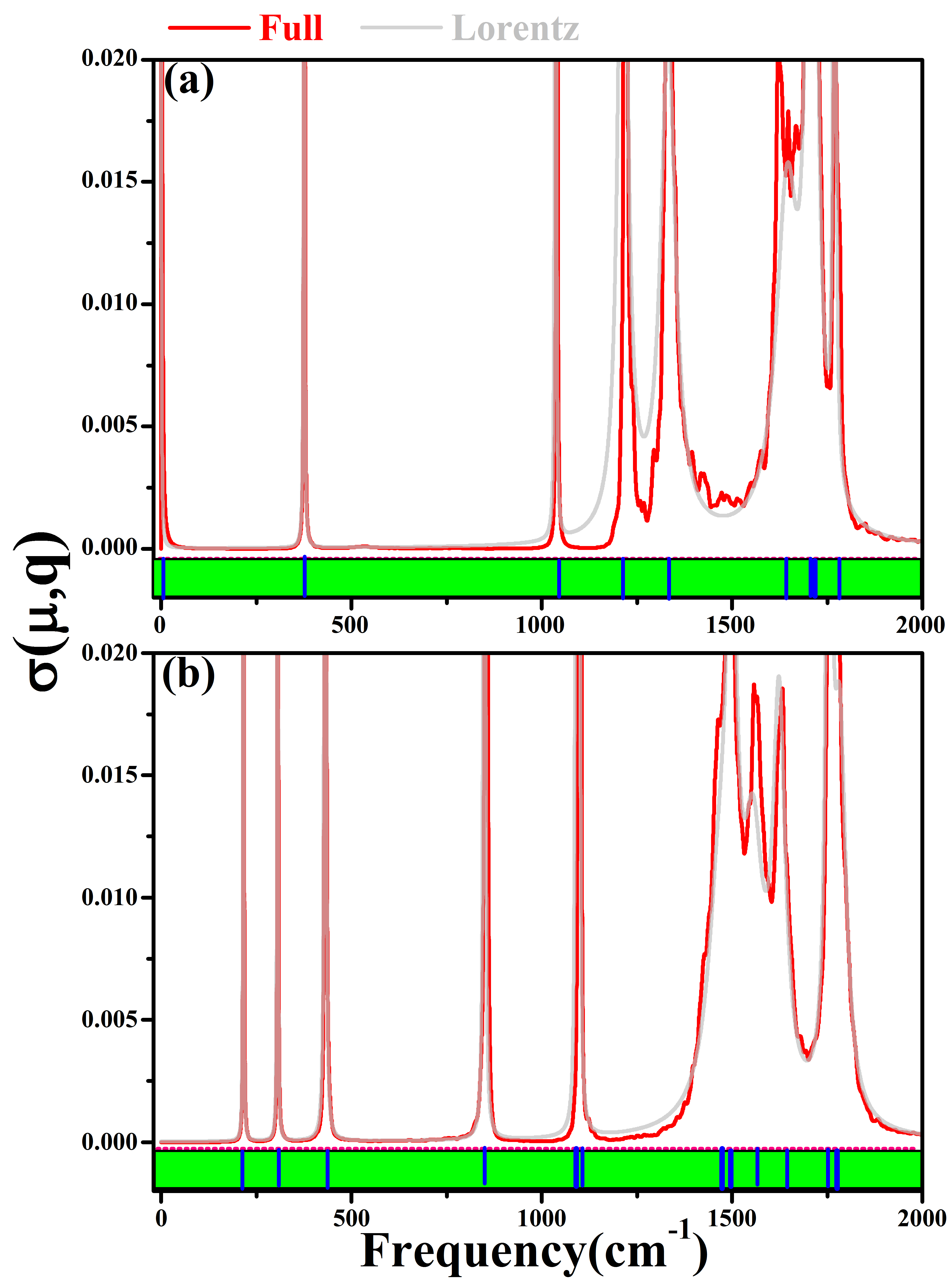}
\caption{(Color online) Phonon spectral function $\sigma(\bq,\Omega)$ of \AlH at 99 GPa (pressure calculated with anharmonic quantum effects), at (a) the 
$\Gamma$ point and (b) the X point.  The red line indicates the result obtained keeping the full energy dependence on the self-energy and the grey line indicates the spectrum calculated in the Lorentzian approximation\cite{29,PhysRevB.97.214101}. The centers of these Lorentzians define the anharmonic phonon frequencies. They are indicated with the blue short vertical lines in the lower panel with green background.}
\label{3.jpg} 
\end{figure}

\section{Results and Discussions}
\label{sec:results}

\subsection{Pressure and crystal structure}

\pstmmc \AlH has a very high symmetry (see Fig. \ref{1.png}) with 8 Al atoms in the corners and 1 Al atom in the center of the cubic unit cell. For each H atom there are 4 nearest Al neighbors at the same distance, while there are 12 equivalent H atom neighbors for each Al atom. Each H atom is located at an interstitial site, in the center of a regular tetrahedron formed by 4 Al atoms. All atomic positions are fixed by symmetry, and, as symmetry is imposed by the SSCHA, internal coordinates of the structure are not affected by quantum effects.

However, the lattice parameter of the cubic structure is subject to quantum effects. 
In fact, as we show in Fig. \ref{1.png} there are strong corrections to the pressure 
of the equation of states if ionic quantum effects are considered. 
For the same lattice parameter, the pressure obtained from the classical
calculation based on the BOES is always about 10 GPa lower than the quantum result
obtained with the SSCHA. This result is rather general among superhydrides, 
as similar quantum corrections on the pressure of about 10 GPa have been 
estimated for H$_3$S and LaH$_{10}$\cite{24,25}. 
Fig. \ref{1.png} can be used conveniently to compare our results
with previous classical calculations\cite{28,19}. For instance, we clearly mark that the classical 
110 GPa and 125 GPa values correspond to 121.2 GPa and 138.1 GPa in the quantum case,
respectively. Consistently, in order to avoid any confusion, 
in the rest of the paper the pressure assigned to harmonic 
calculations will be the the classical one, while the quantum pressure will be assigned to quantum anharmonic calculations.

\subsection{Phonon Spectrum}

The \pstmmc phase of \AlH was observed experimentally above 100 GPa\cite{19}. As shown in Fig. \ref{2.jpg}, 
approximately below this pressure
the system develops phonon instabilities at the X point of the BZ in the classical harmonic
calculation. On the contrary, the anharmonic phonons
obtained diagonalizing $\bDF$ are always stable in the  experimentally relevant pressure range.
Therefore, quantum anharmonic effects play a crucial role in stabilizing the \pstmmc phase of \AlH
around 100 GPa. This phase remains dynamically stable at least down to
70 GPa. This means that even if below 100 GPa \AlH was found in an insulating $P1$ phase, 
the metallic phase may be metastable at lower pressures.

As shown in Fig. \ref{2.jpg}, the anharmonic correction leads to strong changes in the harmonic spectrum
both for the low-energy acoustic and high-energy optical modes. Especially, the phonon frequencies at 
the X point of the BZ are strongly hardened by anharmonicity. Even if the anharmonic hardening of the 
phonon modes at the X point was already anticipated by the early calculations 
in Ref. \cite{28}, the fact that
the anharmonic correction is of the order of the phonon frequency itself questions 
the perturbative approach followed previously. In fact, when using the 
5.933~a\textsubscript{0} lattice parameter, 
which corresponds to 99 GPa if quantum effects are considered, the instabilities apparent in the
harmonic case completely hinder any perturbative approach.

In Fig. \ref{3.jpg} we show the phonon spectral function $\sigma(\bq,\Omega)$ calculated at the $\Gamma$ and X points. These spectral functions can be directly probed by inelastic
x-ray or neutron scattering experiments\cite{PhysRevB.97.214101}. We calculate the spectral
function both keeping the full energy dependence of the phonon self-energy and
within the so-called Lorentzian approximation (see Ref. \cite{PhysRevB.97.214101}
for details). While in the latter case the phonon peaks have, by construction, a Lorentzian lineshape with 
a well-defined linewidth and clear peak position at the $\bbOmega_{\mu}(\bq)$ energies, 
in the former case quasiparticle peaks are not necessarily 
well determined. Despite the large anharmonic correction affecting the
phonon frequencies, all phonon modes keep a well-defined Lorentzian lineshape (see Fig. \ref{3.jpg}),
also for the modes that suffer the largest correction at the X point.
The linewidth of the phonons (half-width at half maximum, HWHM) is very small for the phonon modes below
1100 $cm^{-1}$, less than 1 $cm^{-1}$, while for higher energy modes it is in the range of 
$\sim$ 10 $cm^{-1}$ (see Table \ref{tab:my_label}). It is remarkable that the phonon 
modes derived diagonalizing the free energy Hessian $\bDF$ agree well with the peaks of the 
spectral function (see Table \ref{tab:my_label}), underlining that the phonon modes obtained
in the static limit agree well with the peaks of the dynamical theory and are a valid, for instance,
to study superconducting properties. 

\begin{table}[h]
    \centering
    \caption{The $\Omega_{\mu}(\bq)$ frequencies obtained from the free energy Hessian $\bDF$,  $\bbOmega_{\mu}(\bq)$ frequencies representing the location of the peaks in the spectral function in the Lorentizan approximation, and the anharmonic HWHM linewidth in the latter approximation for the phonon modes at the $\Gamma$ and X points at 99 GPa (pressure calculated including quantum anharmonic effects).}
    \begin{tabular}{c c c c | c c c c }
             \multicolumn{4}{c}{$\Gamma$ point} & \multicolumn{4}{c}{X point} \\
        \hline 
        \hline
            Mode & $\Omega_{\mu}(\bq)$ & $\bbOmega_{\mu}(\bq)$ & $\gamma^{\textrm{anh}}_{\mu}(\bq)$ & Mode & $\Omega_{\mu}(\bq)$ & $\bbOmega_{\mu}(\bq)$ & $\gamma^{\textrm{anh}}_{\mu}(\bq)$\\
        \hline
        1-3    & 0.0 & 0.0 & 0.0 & 1-2 & 213.5 & 216.8 & 0.0\\
        \hline
        4-6    & 378.6 & 376.3 & 0.1 & 3-4 & 305.7 & 306.6 & 0.1\\
        \hline
        7-9    & 1068.7 & 1036.9 & 0.2 & 5-6 & 428.0 & 431.8 & 0.6\\
        \hline
        10-12    & 1254.5 & 1214.2 & 13.3 & 7-8 & 880.2 & 849.9 & 0.9\\
        \hline
        13-15    & 1369.5 & 1332.9 & 23.3 & 9-10 &  1122.9 & 1098.7 & 0.8\\
        \hline
        16-18    & 1645.3 & 1645.3 & 34.8 & 11-12 & 1127.9 & 1091.4 & 0.6\\
        \hline
        19-21    & 1699.7 & 1709.3 & 13.9 & 13-14 & 1495.7 & 1468.2 & 36.7\\
        \hline
        22-23    & 1703.6 & 1712.7 & 4.4 & 15-16 & 1506.7 & 1499.9 & 15.7\\
        \hline
        24      & 1789.4  & 1770.3  & 3.7  & 17-18 & 1572.0 & 1556.2 & 31.7\\
        \hline
             &  &  &  & 19-20 & 1622.9 & 1624.4 & 19.7\\
        \hline
         &  &  &  & 21-22 & 1764.5 & 1755.5 & 6.1\\
        \hline
         &  &  &  & 23-24 & 1773.4 & 1783.8 & 20.3\\
        \hline
        \hline
    \end{tabular}
    \label{tab:my_label}
\end{table}

\begin{figure}[t!]
\includegraphics[width=1.0\columnwidth]{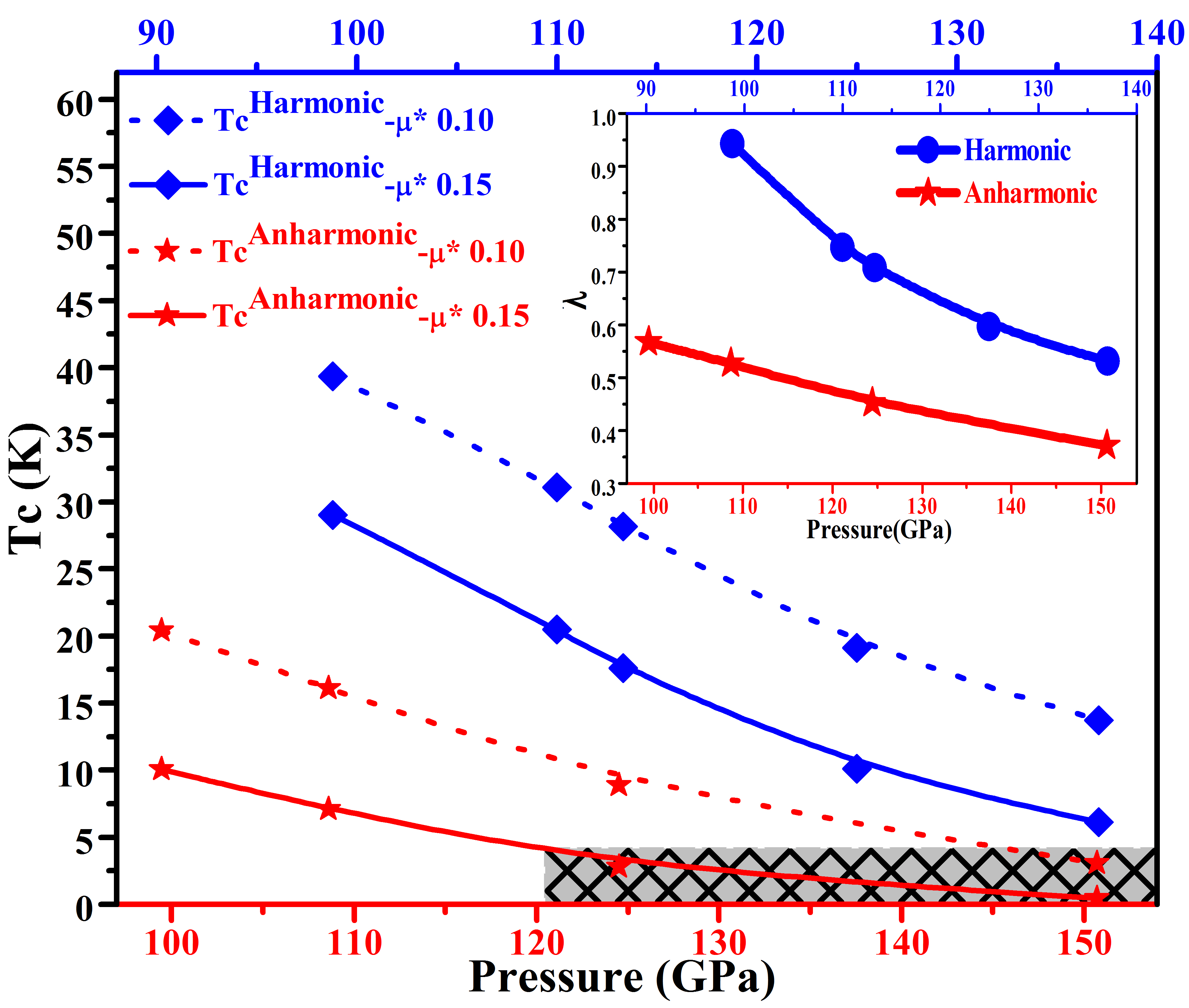}
\caption{(Color online) Superconducting critical temperature $T_c$ and electron-phonon coupling constant $\lambda$ (inset) as a function of pressure in the harmonic approximation (blue lines) and considering anharmonic effects (red lines). Note that a different pressure scale is used for the harmonic and anharmonic calculations, which includes quantum effects for the latter but not the former. Harmonic and anharmonic results aligned vertically are calculated with the same lattice parameter.
$T_c$ calculated with $\mu^*$ = 0.10 and 0.15 is plotted with dotted lines and solid lines respectively.
The grey shadow box marks the pressure region in which 
no superconductivity was found experimentally 
above 4 K\cite{19}.}
\label{4.jpg} 
\end{figure}

\subsection{Superconductivity}

The strong anharmonic renormalization of the phonon spectra has a deep impact on the calculated
superconducting critical temperatures. 
We find that calculations based on the harmonic phonon spectrum 
largely overestimate  $T_c$. We choose typical values for $\mu^*$ 
as 0.10 and 0.15 in the calculations. 
For a given $\mu^*$, $T_c$ decreases monotonically with increasing pressure 
both in the harmonic and anharmonic calculations. 
As shown in Fig. \ref{4.jpg}, for $\mu^*$ = 0.15, the resultant $T_c$ 
values are 7.1, 2.8, and 0.5 K for 109, 125 and 151 GPa, respectively.
The results obtained for the \pstmmc structure using anharmonic phonon frequencies 
agree well with the electrical resistance experiments\cite{19}, 
which reported that there is no superconducting transition above 4 K 
in the 120-164 GPa pressure range.
If the harmonic phonon spectrum is used instead, $T_c$ values above
4 K are predicted even with the largest value of $\mu^*$, completely contradicting the
experimental observation.  Our harmonic calculations are in agreement 
with previous theoretical calculations\cite{19} as 
using McMillan equation\cite{35} with $\mu^*$ = 0.14 $T_c$ is 21.5 K at about 121 GPa.

The anharmonic suppression of $T_c$ is a consequence of the clear drop
of the electron-phonon coupling constant in the anharmonic limit (see Fig. \ref{4.jpg}).
For example, at 109 GPa (pressure calculated with quantum effects), $\lambda$ drops from a value of 
0.95 down to 0.53, a strong suppression, $\lambda$
is practically halved. As a result, given $\mu^*=0.15$, $T_c$ falls from 29 K to 7 K, 
which is equivalent to only 24\% of the harmonic result.
The suppression is similarly impressive for all studied pressures,
whatever the value of $\mu^*$ is. The suppression of superconductivity 
in \pstmmc \AlH is as strong as the one estimated for PdH at ambient 
pressure\cite{26} and PtH at high pressures\cite{22}, which
crystallize in high symmetry phases with H atoms in interstitial sites. 
This suggests that anharmonic suppression of $T_c$ in metallic hydrides with 
isolated H atoms in interstitial sites may be rather common.

\begin{figure}[t]
\includegraphics[width=1.0\columnwidth]{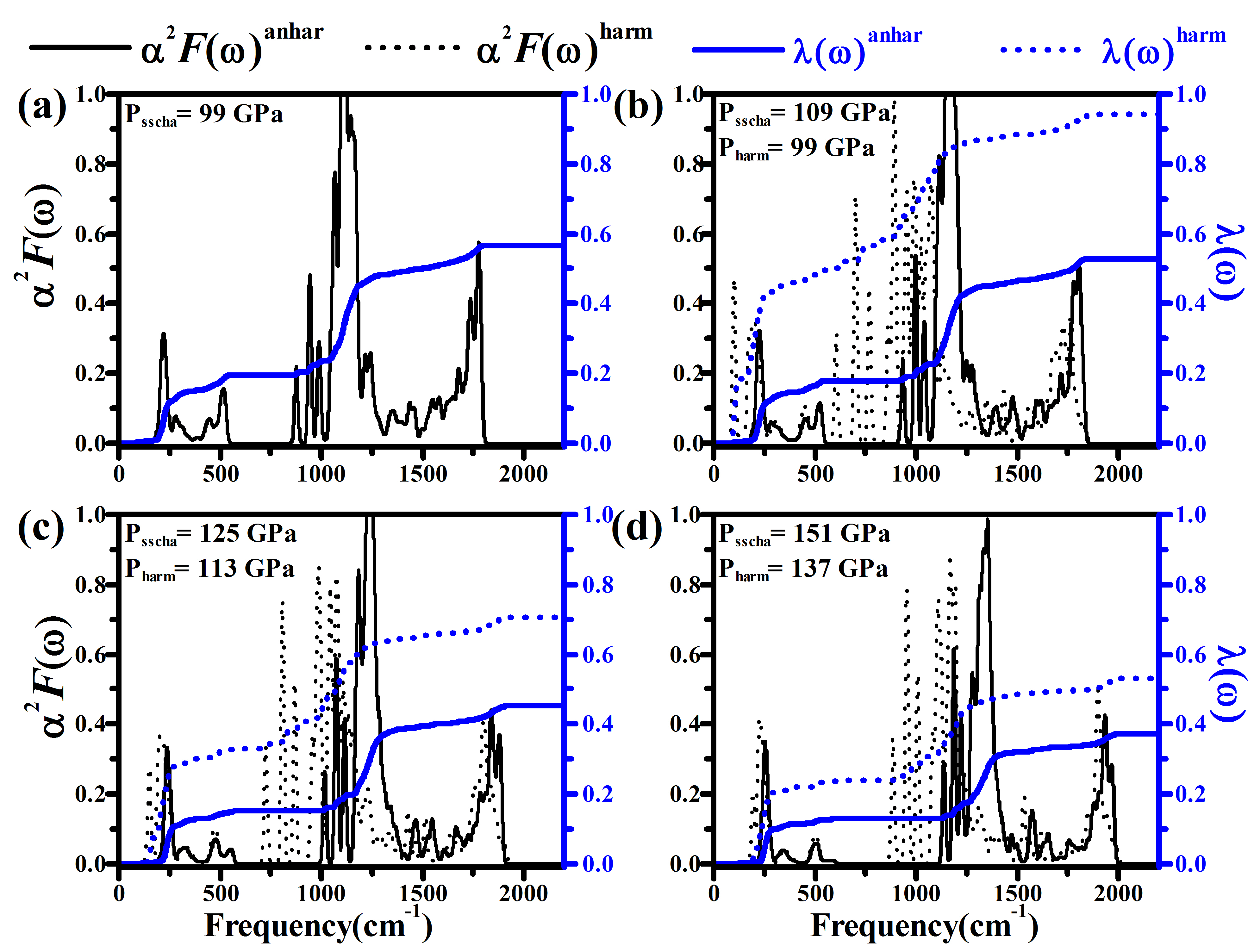}
\caption{Anharmonic spectral function $\alpha^2F(\omega)$ (black solid lines) and integrated
electron-phonon coupling constant $\lambda(\omega)$ (blue solid lines) 
at four different quantum pressures: (a) 99 GPa, (b) 109 GPa, (c) 125 GPa,
and (d) 151 GPa, respectively. The harmonic 
results obtained 
are also shown with dotted lines using the same colors  for comparison at three different classical pressures: (b) 99 GPa, (c) 113 GPa and (d) 137 GPa.
The harmonic and anharmonic results in each panel are obtained with the same 
lattice parameter. }
\label{5.jpg} 
\end{figure}

\begin{figure}[t]
\includegraphics[width=1.0\linewidth]{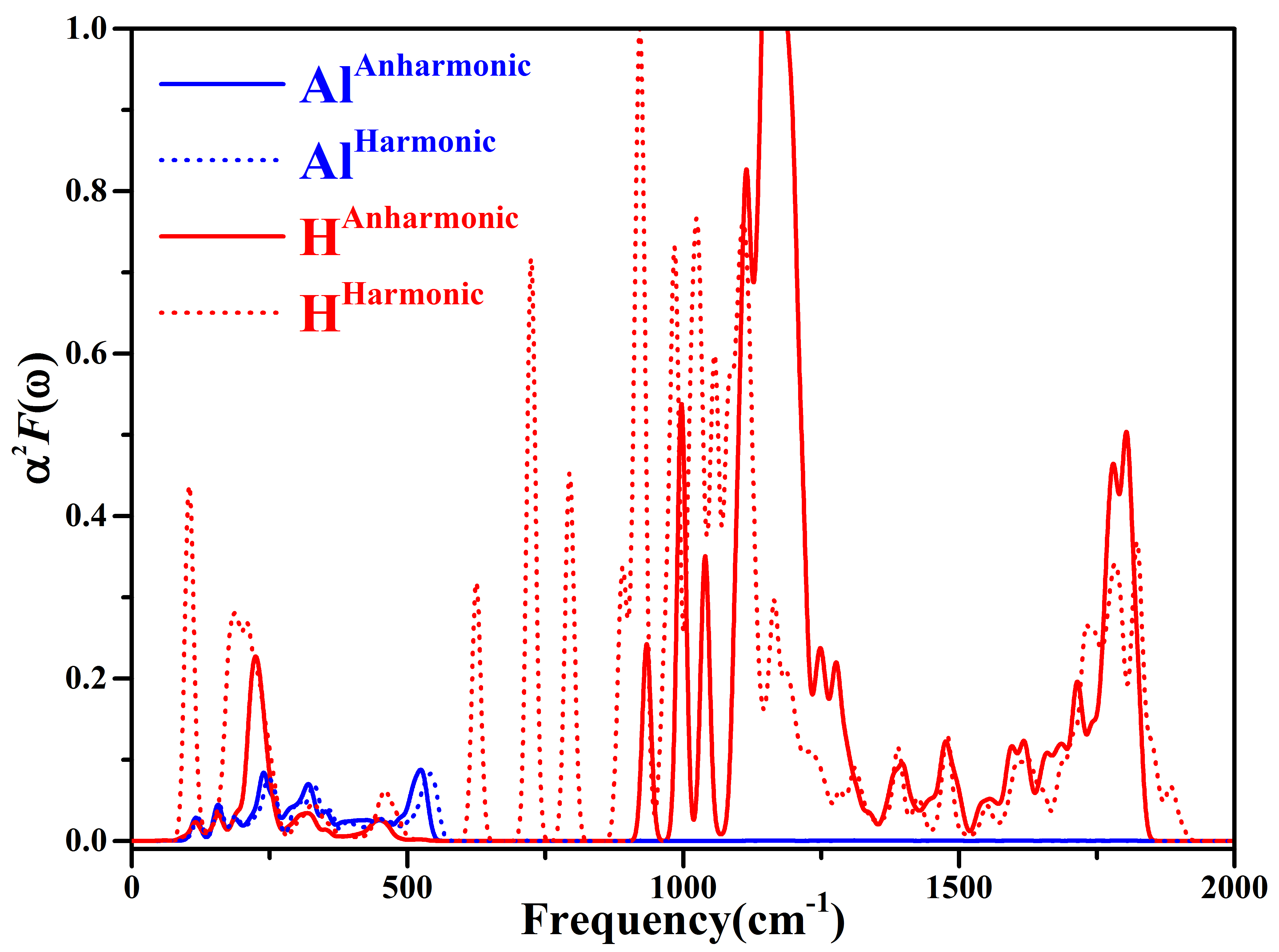}
\caption{The projected $\alpha^2F(\omega)$  onto Al and H at the harmonic 
and anharmonic calculated with the same lattice parameter 5.884 a$_0$, which corresponds in the quantum anharmonic case to 109 GPa. The harmonic results are shown with dotted lines using same colors for comparison.
}
\label{6.jpg} 
\end{figure}

The Eliashberg spectral function $\alpha^2F(\omega)$ and its integral $\lambda(\omega)$, 
both for anharmonic and harmonic cases, are shown in Fig. \ref{5.jpg}. 
It is evident in the figure that while in the anharmonic case the contribution of the low-energy
acoustic modes to $\lambda$ is around 0.2 at all pressures, in the harmonic case it is 
much larger and it is strongly suppressed by pressure. The reason is that in the
harmonic approximation at low pressures there is a significant mixing between H and Al character
in the polarization vectors of the acoustic modes. Pressure lifts the frequencies
of the H-character modes, reducing effectively the mixing. Anharmonicity also suppresses
partly this mixing, and consequently acoustic modes have a weak contribution to $\lambda$. 
This can be seen in Fig. \ref{6.jpg}, where it is evident that
the hydrogen contribution to $\alpha^2F(\omega)$ is suppressed by anharmonicity
at low energies. The contribution of each particular atom to the Eliashberg
function can be obtained by writing $\alpha^2F(\omega)=\sum_{\bar{a}\bar{b}} \alpha^2F_{\bar{a}\bar{b}}(\omega)$ ($\alpha^2F_{\bar{a}\bar{b}}(\omega)$ can be trivially obtained from Eqs. \eqref{eq:eliashberg} and \eqref{eq:elphgamma}). The partial contributions of Al and H
in Fig. \ref{6.jpg} are obtained by summing the contributions in $\alpha^2F_{\bar{a}\bar{b}}(\omega)$ of only Al or H atoms, respectively.
The large peaks in the harmonic $\alpha^2F(\omega)$ in the 550-1000 $cm^{-1}$
frequency range have a very large contribution to $\lambda$ and come from the softened modes
in the vicinity of the X point (see Fig. \ref{2.jpg}). This 
suggests that the softened phonon frequencies at the X point
give a large contribution to the electron-phonon coupling. 
As discussed above, anharmonicity increases these frequencies and, consequently,
shifts these peaks to higher energies.
As the contribution to $\lambda$ of a given mode goes as $\lambda_{\mu}(\bq)=\gamma_{\mu}(\bq)/(N(0)\pi\omega^{2}_{\mu}(\bq))$, 
$\lambda$ is strongly suppressed by anharmonicity. 
Due to the small renormalization of other modes beyond the X point, 
it is reasonable to assume that the bulk of the anharmonic correction
to $\lambda$ and $T_c$ concentrates in the vicinity of X.


\begin{figure*}
\includegraphics[scale=2]{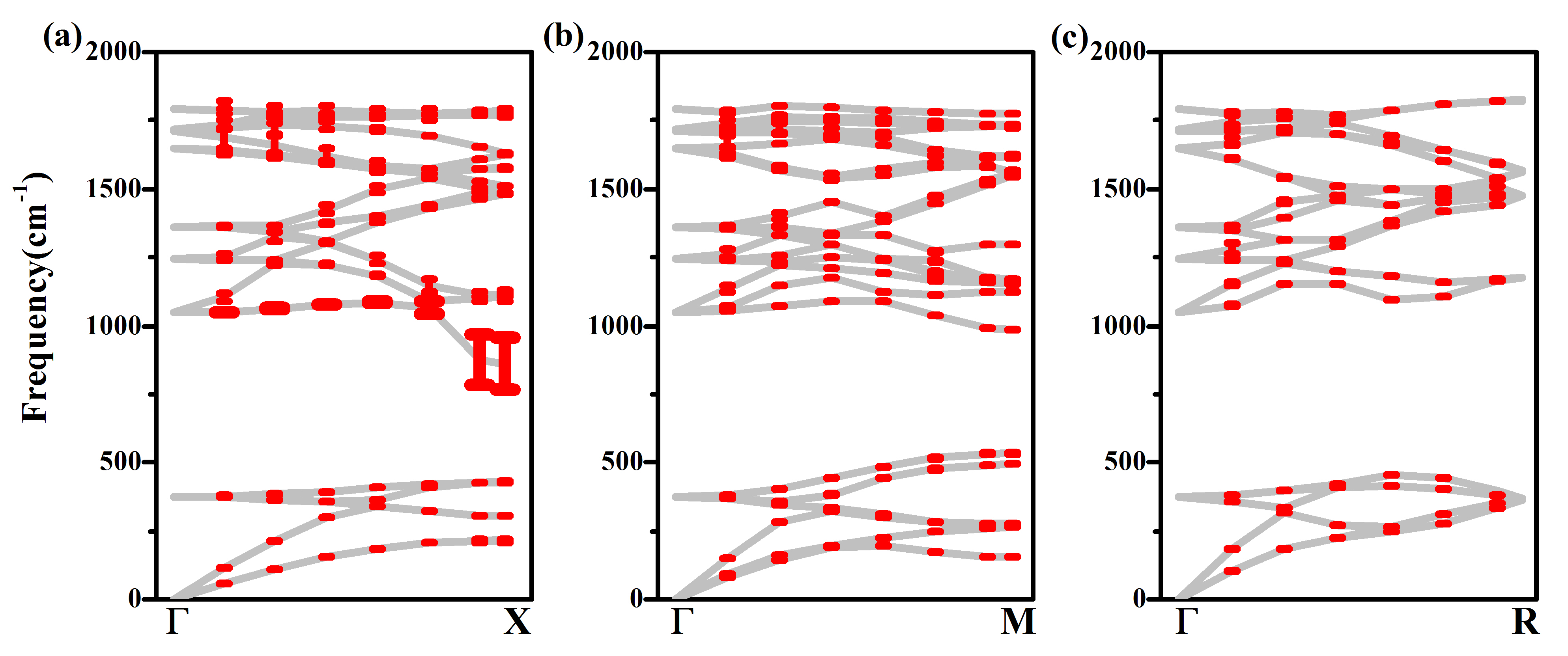}
\caption{(Color online) The linewidth associated to the electron-phonon interaction
calculated with the spectra obtained from $\bDF$ for the cubic 
high-symmetry \pstmmc phase of \AlH at 99 GPa (pressure calculated including quantum effects).
(a) $\Gamma-X$ path, (b) $\Gamma-M$ path, and (c) $\Gamma-R$ path. 
The phonon linewidth is indicated by the size of the red error bars.}
\label{7.jpg} 
\end{figure*}


The fact that the bulk electron-phonon interaction is concentrated around
the X point for the lowest-energy H-character mode is evident when plotting the
$\gamma_{\mu}(\bq)$ linewidth associated to the electron-phonon coupling. 
In Fig. \ref{7.jpg} we show the linewidth 
calculated following Eq. \eqref{eq:elphgamma} for the $\Gamma-X$, $\Gamma-M$, 
and $\Gamma-R$ paths at 99 GPa (pressure calculated with quantum anharmonic effects).
As depicted, the linewidth associated to the lowest-energy H-character mode
largely outweights the contribution of all other modes, underlining that 
this is the mode that contributes the most to $\lambda$. This 
also naturally explains the large anharmonic correction to $T_c$, as
the frequency of this mode is strongly enhanced by anharmonicity.

In our calculations, the electron-phonon contribution to the phonon linewidth of this mode is 90 $cm^{-1}$ 
at 99 GPa and 89.9 $cm^{-1}$ for 109 GPa, respectively (pressures evaluated including quantum anharmonic effects).
This shows that the electron-phonon matrix elements are weakly pressure dependent
in this system. For this mode, the electron-phonon contribution to the linewidth 
is clearly much larger than the anharmonic contribution, which is only 0.9 $cm^{-1}$
(see modes 7-8 at the X point in Table \ref{tab:my_label}).
The electron-phonon contribution to the linewidth is not so large for all the other modes too,
and is comparable (if not smaller) than the anharmonic contribution. 
Interestingly, even if $\gamma_{\mu}(\bq)$ does not depend on the 
phonon frequencies, the electron-phonon linewidth of the strongly renormalized 
phonon mode at the X point is smaller in the harmonic approximation for the same lattice parameter: 78 $cm^{-1}$. The explanation to this is the 
change in the polarization vectors imposed by anharmonicity, which is consistent
with the reduction of the H character of the acoustic modes described above. 
This means that the effect of anharmonicity on 
the electron-phonon coupling constant and $T_c$ cannot be simply reduced to a renormalization
of the phonon frequencies, as it affects them also through a change of the polarization vectors.
Since these effects are not trivial and their impact on $T_c$ cannot be easily anticipated, this motivates
the necessity of performing full nonperturbative anharmonic calculation on 
hydrides to have reliable results on the electron-phonon coupling effects and superconducting properties.

\section{Conclusions}
\label{sec:conclusions}

In summary, in this work we demonstrate clearly that quantum anharmonic effects
are responsible for the absence of superconductivity in \pstmmc  \AlH under high pressure,
confirming the early suggestions\cite{28}. 
We find that the phonon spectra are strongly affected by anharmonic effects, 
which leads the structure to be dynamically stable at lower pressures than expected 
within classical harmonic calculations. 
Anharmonicity reduces the electron-phonon coupling constant 
by no less than 30\% and $T_c$ by at least 59\% in the range of 109-151 GPa.
The bulk of the anharmonic correction, as well as the electron-phonon
interaction, concentrates around the zone border X point.
The \pstmmc remains metastable (because dynamically stable) below
100 GPa, opening the possibility of its synthesis below this pressure.
Our work underlines that superconducting properties of hydrides at high pressure 
can only be properly described by including quantum and anharmonic effects.

\section*{Acknowledgements}

This research was supported by the European Research Council (ERC) 
under the European Unions Horizon 2020 research and innovation programme (grant agreement No. 802533).


\begin{thebibliography}{52}
\expandafter\ifx\csname natexlab\endcsname\relax\def\natexlab#1{#1}\fi
\expandafter\ifx\csname bibnamefont\endcsname\relax
  \def\bibnamefont#1{#1}\fi
\expandafter\ifx\csname bibfnamefont\endcsname\relax
  \def\bibfnamefont#1{#1}\fi
\expandafter\ifx\csname citenamefont\endcsname\relax
  \def\citenamefont#1{#1}\fi
\expandafter\ifx\csname url\endcsname\relax
  \def\url#1{\texttt{#1}}\fi
\expandafter\ifx\csname urlprefix\endcsname\relax\def\urlprefix{URL }\fi
\providecommand{\bibinfo}[2]{#2}
\providecommand{\eprint}[2][]{\url{#2}}

\bibitem[{\citenamefont{Ashcroft}(1968)}]{1}
\bibinfo{author}{\bibfnamefont{N.~W.} \bibnamefont{Ashcroft}},
  \bibinfo{journal}{Phys. Rev. Lett.} \textbf{\bibinfo{volume}{21}},
  \bibinfo{pages}{1748} (\bibinfo{year}{1968}),
  \urlprefix\url{https://link.aps.org/doi/10.1103/PhysRevLett.21.1748}.

\bibitem[{\citenamefont{Flores-Livas et~al.}(2020)\citenamefont{Flores-Livas,
  Boeri, Sanna, Profeta, Arita, and Eremets}}]{FLORESLIVAS20201}
\bibinfo{author}{\bibfnamefont{J.~A.} \bibnamefont{Flores-Livas}},
  \bibinfo{author}{\bibfnamefont{L.}~\bibnamefont{Boeri}},
  \bibinfo{author}{\bibfnamefont{A.}~\bibnamefont{Sanna}},
  \bibinfo{author}{\bibfnamefont{G.}~\bibnamefont{Profeta}},
  \bibinfo{author}{\bibfnamefont{R.}~\bibnamefont{Arita}}, \bibnamefont{and}
  \bibinfo{author}{\bibfnamefont{M.}~\bibnamefont{Eremets}},
  \bibinfo{journal}{Physics Reports} \textbf{\bibinfo{volume}{856}},
  \bibinfo{pages}{1 } (\bibinfo{year}{2020}), ISSN \bibinfo{issn}{0370-1573},
  \bibinfo{note}{a perspective on conventional high-temperature superconductors
  at high pressure: Methods and materials},
  \urlprefix\url{http://www.sciencedirect.com/science/article/pii/S0370157320300363}.

\bibitem[{\citenamefont{Bi et~al.}(2019)\citenamefont{Bi, Zarifi, Terpstra, and
  Zurek}}]{BI2019}
\bibinfo{author}{\bibfnamefont{T.}~\bibnamefont{Bi}},
  \bibinfo{author}{\bibfnamefont{N.}~\bibnamefont{Zarifi}},
  \bibinfo{author}{\bibfnamefont{T.}~\bibnamefont{Terpstra}}, \bibnamefont{and}
  \bibinfo{author}{\bibfnamefont{E.}~\bibnamefont{Zurek}}, in
  \emph{\bibinfo{booktitle}{Reference Module in Chemistry, Molecular Sciences
  and Chemical Engineering}} (\bibinfo{publisher}{Elsevier},
  \bibinfo{year}{2019}), ISBN \bibinfo{isbn}{978-0-12-409547-2},
  \urlprefix\url{http://www.sciencedirect.com/science/article/pii/B9780124095472114350}.

\bibitem[{\citenamefont{Pickard et~al.}(2020)\citenamefont{Pickard, Errea, and
  Eremets}}]{doi:10.1146/annurev-conmatphys-031218-013413}
\bibinfo{author}{\bibfnamefont{C.~J.} \bibnamefont{Pickard}},
  \bibinfo{author}{\bibfnamefont{I.}~\bibnamefont{Errea}}, \bibnamefont{and}
  \bibinfo{author}{\bibfnamefont{M.~I.} \bibnamefont{Eremets}},
  \bibinfo{journal}{Annual Review of Condensed Matter Physics}
  \textbf{\bibinfo{volume}{11}}, \bibinfo{pages}{57} (\bibinfo{year}{2020}),
  \eprint{https://doi.org/10.1146/annurev-conmatphys-031218-013413},
  \urlprefix\url{https://doi.org/10.1146/annurev-conmatphys-031218-013413}.

\bibitem[{\citenamefont{Gao et~al.}(2008)\citenamefont{Gao, Oganov, Bergara,
  Martinez-Canales, Cui, Iitaka, Ma, and Zou}}]{2}
\bibinfo{author}{\bibfnamefont{G.}~\bibnamefont{Gao}},
  \bibinfo{author}{\bibfnamefont{A.~R.} \bibnamefont{Oganov}},
  \bibinfo{author}{\bibfnamefont{A.}~\bibnamefont{Bergara}},
  \bibinfo{author}{\bibfnamefont{M.}~\bibnamefont{Martinez-Canales}},
  \bibinfo{author}{\bibfnamefont{T.}~\bibnamefont{Cui}},
  \bibinfo{author}{\bibfnamefont{T.}~\bibnamefont{Iitaka}},
  \bibinfo{author}{\bibfnamefont{Y.}~\bibnamefont{Ma}}, \bibnamefont{and}
  \bibinfo{author}{\bibfnamefont{G.}~\bibnamefont{Zou}},
  \bibinfo{journal}{Phys. Rev. Lett.} \textbf{\bibinfo{volume}{101}},
  \bibinfo{pages}{107002} (\bibinfo{year}{2008}),
  \urlprefix\url{https://link.aps.org/doi/10.1103/PhysRevLett.101.107002}.

\bibitem[{\citenamefont{Martinez-Canales
  et~al.}(2009)\citenamefont{Martinez-Canales, Oganov, Ma, Yan, Lyakhov, and
  Bergara}}]{3}
\bibinfo{author}{\bibfnamefont{M.}~\bibnamefont{Martinez-Canales}},
  \bibinfo{author}{\bibfnamefont{A.~R.} \bibnamefont{Oganov}},
  \bibinfo{author}{\bibfnamefont{Y.}~\bibnamefont{Ma}},
  \bibinfo{author}{\bibfnamefont{Y.}~\bibnamefont{Yan}},
  \bibinfo{author}{\bibfnamefont{A.~O.} \bibnamefont{Lyakhov}},
  \bibnamefont{and} \bibinfo{author}{\bibfnamefont{A.}~\bibnamefont{Bergara}},
  \bibinfo{journal}{Phys. Rev. Lett.} \textbf{\bibinfo{volume}{102}},
  \bibinfo{pages}{087005} (\bibinfo{year}{2009}),
  \urlprefix\url{https://link.aps.org/doi/10.1103/PhysRevLett.102.087005}.

\bibitem[{\citenamefont{Gao et~al.}(2010)\citenamefont{Gao, Oganov, Li, Li,
  Wang, Cui, Ma, Bergara, Lyakhov, Iitaka et~al.}}]{4}
\bibinfo{author}{\bibfnamefont{G.}~\bibnamefont{Gao}},
  \bibinfo{author}{\bibfnamefont{A.~R.} \bibnamefont{Oganov}},
  \bibinfo{author}{\bibfnamefont{P.}~\bibnamefont{Li}},
  \bibinfo{author}{\bibfnamefont{Z.}~\bibnamefont{Li}},
  \bibinfo{author}{\bibfnamefont{H.}~\bibnamefont{Wang}},
  \bibinfo{author}{\bibfnamefont{T.}~\bibnamefont{Cui}},
  \bibinfo{author}{\bibfnamefont{Y.}~\bibnamefont{Ma}},
  \bibinfo{author}{\bibfnamefont{A.}~\bibnamefont{Bergara}},
  \bibinfo{author}{\bibfnamefont{A.~O.} \bibnamefont{Lyakhov}},
  \bibinfo{author}{\bibfnamefont{T.}~\bibnamefont{Iitaka}},
  \bibnamefont{et~al.}, \bibinfo{journal}{Proceedings of the National Academy
  of Sciences} \textbf{\bibinfo{volume}{107}}, \bibinfo{pages}{1317}
  (\bibinfo{year}{2010}), ISSN \bibinfo{issn}{0027-8424},
  \eprint{https://www.pnas.org/content/107/4/1317.full.pdf},
  \urlprefix\url{https://www.pnas.org/content/107/4/1317}.

\bibitem[{\citenamefont{Kim et~al.}(2011)\citenamefont{Kim, Scheicher, Pickard,
  Needs, and Ahuja}}]{5}
\bibinfo{author}{\bibfnamefont{D.~Y.} \bibnamefont{Kim}},
  \bibinfo{author}{\bibfnamefont{R.~H.} \bibnamefont{Scheicher}},
  \bibinfo{author}{\bibfnamefont{C.~J.} \bibnamefont{Pickard}},
  \bibinfo{author}{\bibfnamefont{R.~J.} \bibnamefont{Needs}}, \bibnamefont{and}
  \bibinfo{author}{\bibfnamefont{R.}~\bibnamefont{Ahuja}},
  \bibinfo{journal}{Phys. Rev. Lett.} \textbf{\bibinfo{volume}{107}},
  \bibinfo{pages}{117002} (\bibinfo{year}{2011}),
  \urlprefix\url{https://link.aps.org/doi/10.1103/PhysRevLett.107.117002}.

\bibitem[{\citenamefont{Wang et~al.}(2012)\citenamefont{Wang, Tse, Tanaka,
  Iitaka, and Ma}}]{6}
\bibinfo{author}{\bibfnamefont{H.}~\bibnamefont{Wang}},
  \bibinfo{author}{\bibfnamefont{J.~S.} \bibnamefont{Tse}},
  \bibinfo{author}{\bibfnamefont{K.}~\bibnamefont{Tanaka}},
  \bibinfo{author}{\bibfnamefont{T.}~\bibnamefont{Iitaka}}, \bibnamefont{and}
  \bibinfo{author}{\bibfnamefont{Y.}~\bibnamefont{Ma}},
  \bibinfo{journal}{Proceedings of the National Academy of Sciences}
  \textbf{\bibinfo{volume}{109}}, \bibinfo{pages}{6463} (\bibinfo{year}{2012}),
  ISSN \bibinfo{issn}{0027-8424},
  \eprint{https://www.pnas.org/content/109/17/6463.full.pdf},
  \urlprefix\url{https://www.pnas.org/content/109/17/6463}.

\bibitem[{\citenamefont{Lonie et~al.}(2013)\citenamefont{Lonie, Hooper,
  Altintas, and Zurek}}]{7}
\bibinfo{author}{\bibfnamefont{D.~C.} \bibnamefont{Lonie}},
  \bibinfo{author}{\bibfnamefont{J.}~\bibnamefont{Hooper}},
  \bibinfo{author}{\bibfnamefont{B.}~\bibnamefont{Altintas}}, \bibnamefont{and}
  \bibinfo{author}{\bibfnamefont{E.}~\bibnamefont{Zurek}},
  \bibinfo{journal}{Phys. Rev. B} \textbf{\bibinfo{volume}{87}},
  \bibinfo{pages}{054107} (\bibinfo{year}{2013}),
  \urlprefix\url{https://link.aps.org/doi/10.1103/PhysRevB.87.054107}.

\bibitem[{\citenamefont{Duan et~al.}(2014)\citenamefont{Duan, Liu, Tian, Li,
  Huang, Zhao, Yu, Liu, Tian, and Cui}}]{8}
\bibinfo{author}{\bibfnamefont{D.}~\bibnamefont{Duan}},
  \bibinfo{author}{\bibfnamefont{Y.}~\bibnamefont{Liu}},
  \bibinfo{author}{\bibfnamefont{F.}~\bibnamefont{Tian}},
  \bibinfo{author}{\bibfnamefont{D.}~\bibnamefont{Li}},
  \bibinfo{author}{\bibfnamefont{X.}~\bibnamefont{Huang}},
  \bibinfo{author}{\bibfnamefont{Z.}~\bibnamefont{Zhao}},
  \bibinfo{author}{\bibfnamefont{H.}~\bibnamefont{Yu}},
  \bibinfo{author}{\bibfnamefont{B.}~\bibnamefont{Liu}},
  \bibinfo{author}{\bibfnamefont{W.}~\bibnamefont{Tian}}, \bibnamefont{and}
  \bibinfo{author}{\bibfnamefont{T.}~\bibnamefont{Cui}},
  \bibinfo{journal}{Scientific Reports} \textbf{\bibinfo{volume}{4}},
  \bibinfo{pages}{6968} (\bibinfo{year}{2014}), ISSN \bibinfo{issn}{2045-2322},
  \urlprefix\url{https://doi.org/10.1038/srep06968}.

\bibitem[{\citenamefont{Duan et~al.}(2015)\citenamefont{Duan, Huang, Tian, Li,
  Yu, Liu, Ma, Liu, and Cui}}]{9}
\bibinfo{author}{\bibfnamefont{D.}~\bibnamefont{Duan}},
  \bibinfo{author}{\bibfnamefont{X.}~\bibnamefont{Huang}},
  \bibinfo{author}{\bibfnamefont{F.}~\bibnamefont{Tian}},
  \bibinfo{author}{\bibfnamefont{D.}~\bibnamefont{Li}},
  \bibinfo{author}{\bibfnamefont{H.}~\bibnamefont{Yu}},
  \bibinfo{author}{\bibfnamefont{Y.}~\bibnamefont{Liu}},
  \bibinfo{author}{\bibfnamefont{Y.}~\bibnamefont{Ma}},
  \bibinfo{author}{\bibfnamefont{B.}~\bibnamefont{Liu}}, \bibnamefont{and}
  \bibinfo{author}{\bibfnamefont{T.}~\bibnamefont{Cui}},
  \bibinfo{journal}{Phys. Rev. B} \textbf{\bibinfo{volume}{91}},
  \bibinfo{pages}{180502} (\bibinfo{year}{2015}),
  \urlprefix\url{https://link.aps.org/doi/10.1103/PhysRevB.91.180502}.

\bibitem[{\citenamefont{Li et~al.}(2015)\citenamefont{Li, Hao, Liu, Tse, Wang,
  and Ma}}]{10}
\bibinfo{author}{\bibfnamefont{Y.}~\bibnamefont{Li}},
  \bibinfo{author}{\bibfnamefont{J.}~\bibnamefont{Hao}},
  \bibinfo{author}{\bibfnamefont{H.}~\bibnamefont{Liu}},
  \bibinfo{author}{\bibfnamefont{J.~S.} \bibnamefont{Tse}},
  \bibinfo{author}{\bibfnamefont{Y.}~\bibnamefont{Wang}}, \bibnamefont{and}
  \bibinfo{author}{\bibfnamefont{Y.}~\bibnamefont{Ma}},
  \bibinfo{journal}{Scientific Reports} \textbf{\bibinfo{volume}{5}},
  \bibinfo{pages}{9948} (\bibinfo{year}{2015}), ISSN \bibinfo{issn}{2045-2322},
  \urlprefix\url{https://doi.org/10.1038/srep09948}.

\bibitem[{\citenamefont{Zhang et~al.}(2015)\citenamefont{Zhang, Wang, Zhang,
  Liu, Zhong, Song, Yang, Zhang, and Ma}}]{11}
\bibinfo{author}{\bibfnamefont{S.}~\bibnamefont{Zhang}},
  \bibinfo{author}{\bibfnamefont{Y.}~\bibnamefont{Wang}},
  \bibinfo{author}{\bibfnamefont{J.}~\bibnamefont{Zhang}},
  \bibinfo{author}{\bibfnamefont{H.}~\bibnamefont{Liu}},
  \bibinfo{author}{\bibfnamefont{X.}~\bibnamefont{Zhong}},
  \bibinfo{author}{\bibfnamefont{H.-F.} \bibnamefont{Song}},
  \bibinfo{author}{\bibfnamefont{G.}~\bibnamefont{Yang}},
  \bibinfo{author}{\bibfnamefont{L.}~\bibnamefont{Zhang}}, \bibnamefont{and}
  \bibinfo{author}{\bibfnamefont{Y.}~\bibnamefont{Ma}},
  \bibinfo{journal}{Scientific Reports} \textbf{\bibinfo{volume}{5}},
  \bibinfo{pages}{15433} (\bibinfo{year}{2015}), ISSN
  \bibinfo{issn}{2045-2322}, \urlprefix\url{https://doi.org/10.1038/srep15433}.

\bibitem[{\citenamefont{Liu et~al.}(2016)\citenamefont{Liu, Li, Gao, Tse, and
  Naumov}}]{12}
\bibinfo{author}{\bibfnamefont{H.}~\bibnamefont{Liu}},
  \bibinfo{author}{\bibfnamefont{Y.}~\bibnamefont{Li}},
  \bibinfo{author}{\bibfnamefont{G.}~\bibnamefont{Gao}},
  \bibinfo{author}{\bibfnamefont{J.~S.} \bibnamefont{Tse}}, \bibnamefont{and}
  \bibinfo{author}{\bibfnamefont{I.~I.} \bibnamefont{Naumov}},
  \bibinfo{journal}{The Journal of Physical Chemistry C}
  \textbf{\bibinfo{volume}{120}}, \bibinfo{pages}{3458} (\bibinfo{year}{2016}),
  \eprint{https://doi.org/10.1021/acs.jpcc.5b12009},
  \urlprefix\url{https://doi.org/10.1021/acs.jpcc.5b12009}.

\bibitem[{\citenamefont{Zhong et~al.}(2016)\citenamefont{Zhong, Wang, Zhang,
  Liu, Zhang, Song, Yang, Zhang, and Ma}}]{13}
\bibinfo{author}{\bibfnamefont{X.}~\bibnamefont{Zhong}},
  \bibinfo{author}{\bibfnamefont{H.}~\bibnamefont{Wang}},
  \bibinfo{author}{\bibfnamefont{J.}~\bibnamefont{Zhang}},
  \bibinfo{author}{\bibfnamefont{H.}~\bibnamefont{Liu}},
  \bibinfo{author}{\bibfnamefont{S.}~\bibnamefont{Zhang}},
  \bibinfo{author}{\bibfnamefont{H.-F.} \bibnamefont{Song}},
  \bibinfo{author}{\bibfnamefont{G.}~\bibnamefont{Yang}},
  \bibinfo{author}{\bibfnamefont{L.}~\bibnamefont{Zhang}}, \bibnamefont{and}
  \bibinfo{author}{\bibfnamefont{Y.}~\bibnamefont{Ma}}, \bibinfo{journal}{Phys.
  Rev. Lett.} \textbf{\bibinfo{volume}{116}}, \bibinfo{pages}{057002}
  (\bibinfo{year}{2016}),
  \urlprefix\url{https://link.aps.org/doi/10.1103/PhysRevLett.116.057002}.

\bibitem[{\citenamefont{Peng et~al.}(2017)\citenamefont{Peng, Sun, Pickard,
  Needs, Wu, and Ma}}]{14}
\bibinfo{author}{\bibfnamefont{F.}~\bibnamefont{Peng}},
  \bibinfo{author}{\bibfnamefont{Y.}~\bibnamefont{Sun}},
  \bibinfo{author}{\bibfnamefont{C.~J.} \bibnamefont{Pickard}},
  \bibinfo{author}{\bibfnamefont{R.~J.} \bibnamefont{Needs}},
  \bibinfo{author}{\bibfnamefont{Q.}~\bibnamefont{Wu}}, \bibnamefont{and}
  \bibinfo{author}{\bibfnamefont{Y.}~\bibnamefont{Ma}}, \bibinfo{journal}{Phys.
  Rev. Lett.} \textbf{\bibinfo{volume}{119}}, \bibinfo{pages}{107001}
  (\bibinfo{year}{2017}),
  \urlprefix\url{https://link.aps.org/doi/10.1103/PhysRevLett.119.107001}.

\bibitem[{\citenamefont{Liu et~al.}(2017)\citenamefont{Liu, Naumov, Hoffmann,
  Ashcroft, and Hemley}}]{15}
\bibinfo{author}{\bibfnamefont{H.}~\bibnamefont{Liu}},
  \bibinfo{author}{\bibfnamefont{I.~I.} \bibnamefont{Naumov}},
  \bibinfo{author}{\bibfnamefont{R.}~\bibnamefont{Hoffmann}},
  \bibinfo{author}{\bibfnamefont{N.~W.} \bibnamefont{Ashcroft}},
  \bibnamefont{and} \bibinfo{author}{\bibfnamefont{R.~J.}
  \bibnamefont{Hemley}}, \bibinfo{journal}{Proceedings of the National Academy
  of Sciences} \textbf{\bibinfo{volume}{114}}, \bibinfo{pages}{6990}
  (\bibinfo{year}{2017}), ISSN \bibinfo{issn}{0027-8424},
  \eprint{https://www.pnas.org/content/114/27/6990.full.pdf},
  \urlprefix\url{https://www.pnas.org/content/114/27/6990}.

\bibitem[{\citenamefont{Sun et~al.}(2019)\citenamefont{Sun, Lv, Xie, Liu, and
  Ma}}]{16}
\bibinfo{author}{\bibfnamefont{Y.}~\bibnamefont{Sun}},
  \bibinfo{author}{\bibfnamefont{J.}~\bibnamefont{Lv}},
  \bibinfo{author}{\bibfnamefont{Y.}~\bibnamefont{Xie}},
  \bibinfo{author}{\bibfnamefont{H.}~\bibnamefont{Liu}}, \bibnamefont{and}
  \bibinfo{author}{\bibfnamefont{Y.}~\bibnamefont{Ma}}, \bibinfo{journal}{Phys.
  Rev. Lett.} \textbf{\bibinfo{volume}{123}}, \bibinfo{pages}{097001}
  (\bibinfo{year}{2019}),
  \urlprefix\url{https://link.aps.org/doi/10.1103/PhysRevLett.123.097001}.

\bibitem[{\citenamefont{Cui et~al.}(2020)\citenamefont{Cui, Bi, Shi, Li, Liu,
  Zurek, and Hemley}}]{17}
\bibinfo{author}{\bibfnamefont{W.}~\bibnamefont{Cui}},
  \bibinfo{author}{\bibfnamefont{T.}~\bibnamefont{Bi}},
  \bibinfo{author}{\bibfnamefont{J.}~\bibnamefont{Shi}},
  \bibinfo{author}{\bibfnamefont{Y.}~\bibnamefont{Li}},
  \bibinfo{author}{\bibfnamefont{H.}~\bibnamefont{Liu}},
  \bibinfo{author}{\bibfnamefont{E.}~\bibnamefont{Zurek}}, \bibnamefont{and}
  \bibinfo{author}{\bibfnamefont{R.~J.} \bibnamefont{Hemley}},
  \bibinfo{journal}{Phys. Rev. B} \textbf{\bibinfo{volume}{101}},
  \bibinfo{pages}{134504} (\bibinfo{year}{2020}),
  \urlprefix\url{https://link.aps.org/doi/10.1103/PhysRevB.101.134504}.

\bibitem[{\citenamefont{Sun et~al.}(2020)\citenamefont{Sun, Tian, Jiang, Li,
  Li, Iitaka, Zhong, and Xie}}]{18}
\bibinfo{author}{\bibfnamefont{Y.}~\bibnamefont{Sun}},
  \bibinfo{author}{\bibfnamefont{Y.}~\bibnamefont{Tian}},
  \bibinfo{author}{\bibfnamefont{B.}~\bibnamefont{Jiang}},
  \bibinfo{author}{\bibfnamefont{X.}~\bibnamefont{Li}},
  \bibinfo{author}{\bibfnamefont{H.}~\bibnamefont{Li}},
  \bibinfo{author}{\bibfnamefont{T.}~\bibnamefont{Iitaka}},
  \bibinfo{author}{\bibfnamefont{X.}~\bibnamefont{Zhong}}, \bibnamefont{and}
  \bibinfo{author}{\bibfnamefont{Y.}~\bibnamefont{Xie}},
  \bibinfo{journal}{Phys. Rev. B} \textbf{\bibinfo{volume}{101}},
  \bibinfo{pages}{174102} (\bibinfo{year}{2020}),
  \urlprefix\url{https://link.aps.org/doi/10.1103/PhysRevB.101.174102}.

\bibitem[{\citenamefont{Borinaga et~al.}(2016)\citenamefont{Borinaga, Errea,
  Calandra, Mauri, and Bergara}}]{H:Borinaga_PRB_2016}
\bibinfo{author}{\bibfnamefont{M.}~\bibnamefont{Borinaga}},
  \bibinfo{author}{\bibfnamefont{I.}~\bibnamefont{Errea}},
  \bibinfo{author}{\bibfnamefont{M.}~\bibnamefont{Calandra}},
  \bibinfo{author}{\bibfnamefont{F.}~\bibnamefont{Mauri}}, \bibnamefont{and}
  \bibinfo{author}{\bibfnamefont{A.}~\bibnamefont{Bergara}},
  \bibinfo{journal}{Phys. Rev. B} \textbf{\bibinfo{volume}{93}},
  \bibinfo{pages}{174308} (\bibinfo{year}{2016}),
  \urlprefix\url{http://link.aps.org/doi/10.1103/PhysRevB.93.174308}.

\bibitem[{\citenamefont{Borinaga et~al.}(2018)\citenamefont{Borinaga,
  Iba{\~n}ez-Azpiroz, Bergara, and Errea}}]{borinaga2018strong}
\bibinfo{author}{\bibfnamefont{M.}~\bibnamefont{Borinaga}},
  \bibinfo{author}{\bibfnamefont{J.}~\bibnamefont{Iba{\~n}ez-Azpiroz}},
  \bibinfo{author}{\bibfnamefont{A.}~\bibnamefont{Bergara}}, \bibnamefont{and}
  \bibinfo{author}{\bibfnamefont{I.}~\bibnamefont{Errea}},
  \bibinfo{journal}{Physical review letters} \textbf{\bibinfo{volume}{120}},
  \bibinfo{pages}{057402} (\bibinfo{year}{2018}).

\bibitem[{\citenamefont{Dias and Silvera}(2017)}]{Dias_hydrogen_Science2017}
\bibinfo{author}{\bibfnamefont{R.~P.} \bibnamefont{Dias}} \bibnamefont{and}
  \bibinfo{author}{\bibfnamefont{I.~F.} \bibnamefont{Silvera}},
  \bibinfo{journal}{Science}  (\bibinfo{year}{2017}), ISSN
  \bibinfo{issn}{0036-8075}.

\bibitem[{\citenamefont{Drozdov et~al.}(2015)\citenamefont{Drozdov, Eremets,
  Troyan, Ksenofontov, and Shylin}}]{DrozdovEremets_Nature2015}
\bibinfo{author}{\bibfnamefont{A.~P.} \bibnamefont{Drozdov}},
  \bibinfo{author}{\bibfnamefont{M.~I.} \bibnamefont{Eremets}},
  \bibinfo{author}{\bibfnamefont{I.~A.} \bibnamefont{Troyan}},
  \bibinfo{author}{\bibfnamefont{V.}~\bibnamefont{Ksenofontov}},
  \bibnamefont{and} \bibinfo{author}{\bibfnamefont{S.~I.}
  \bibnamefont{Shylin}}, \bibinfo{journal}{Nature}
  \textbf{\bibinfo{volume}{525}}, \bibinfo{pages}{73} (\bibinfo{year}{2015}).

\bibitem[{\citenamefont{Somayazulu et~al.}(2019)\citenamefont{Somayazulu,
  Ahart, Mishra, Geballe, Baldini, Meng, Struzhkin, and
  Hemley}}]{Hemley-LaH10_PRL_2019}
\bibinfo{author}{\bibfnamefont{M.}~\bibnamefont{Somayazulu}},
  \bibinfo{author}{\bibfnamefont{M.}~\bibnamefont{Ahart}},
  \bibinfo{author}{\bibfnamefont{A.~K.} \bibnamefont{Mishra}},
  \bibinfo{author}{\bibfnamefont{Z.~M.} \bibnamefont{Geballe}},
  \bibinfo{author}{\bibfnamefont{M.}~\bibnamefont{Baldini}},
  \bibinfo{author}{\bibfnamefont{Y.}~\bibnamefont{Meng}},
  \bibinfo{author}{\bibfnamefont{V.~V.} \bibnamefont{Struzhkin}},
  \bibnamefont{and} \bibinfo{author}{\bibfnamefont{R.~J.}
  \bibnamefont{Hemley}}, \bibinfo{journal}{Phys. Rev. Lett.}
  \textbf{\bibinfo{volume}{122}}, \bibinfo{pages}{027001}
  (\bibinfo{year}{2019}),
  \urlprefix\url{https://link.aps.org/doi/10.1103/PhysRevLett.122.027001}.

\bibitem[{\citenamefont{Drozdov et~al.}(2019)\citenamefont{Drozdov, Kong,
  Minkov, Besedin, Kuzovnikov, Mozaffari, Balicas, Balakirev, Graf, Prakapenka
  et~al.}}]{Nature_LaH_Eremets_2019}
\bibinfo{author}{\bibfnamefont{A.~P.} \bibnamefont{Drozdov}},
  \bibinfo{author}{\bibfnamefont{P.~P.} \bibnamefont{Kong}},
  \bibinfo{author}{\bibfnamefont{V.~S.} \bibnamefont{Minkov}},
  \bibinfo{author}{\bibfnamefont{S.~P.} \bibnamefont{Besedin}},
  \bibinfo{author}{\bibfnamefont{M.~A.} \bibnamefont{Kuzovnikov}},
  \bibinfo{author}{\bibfnamefont{S.}~\bibnamefont{Mozaffari}},
  \bibinfo{author}{\bibfnamefont{L.}~\bibnamefont{Balicas}},
  \bibinfo{author}{\bibfnamefont{F.~F.} \bibnamefont{Balakirev}},
  \bibinfo{author}{\bibfnamefont{D.~E.} \bibnamefont{Graf}},
  \bibinfo{author}{\bibfnamefont{V.~B.} \bibnamefont{Prakapenka}},
  \bibnamefont{et~al.}, \bibinfo{journal}{Nature}
  \textbf{\bibinfo{volume}{569}}, \bibinfo{pages}{528} (\bibinfo{year}{2019}),
  \urlprefix\url{https://doi.org/10.1038/s41586-019-1201-8}.

\bibitem[{\citenamefont{Troyan et~al.}(2020)\citenamefont{Troyan, Semenok,
  Kvashnin, Sadakov, Sobolevskiy, Pudalov, Ivanova, Prakapenka, Greenberg,
  Gavriliuk et~al.}}]{troyan2020anomalous}
\bibinfo{author}{\bibfnamefont{I.~A.} \bibnamefont{Troyan}},
  \bibinfo{author}{\bibfnamefont{D.~V.} \bibnamefont{Semenok}},
  \bibinfo{author}{\bibfnamefont{A.~G.} \bibnamefont{Kvashnin}},
  \bibinfo{author}{\bibfnamefont{A.~V.} \bibnamefont{Sadakov}},
  \bibinfo{author}{\bibfnamefont{O.~A.} \bibnamefont{Sobolevskiy}},
  \bibinfo{author}{\bibfnamefont{V.~M.} \bibnamefont{Pudalov}},
  \bibinfo{author}{\bibfnamefont{A.~G.} \bibnamefont{Ivanova}},
  \bibinfo{author}{\bibfnamefont{V.~B.} \bibnamefont{Prakapenka}},
  \bibinfo{author}{\bibfnamefont{E.}~\bibnamefont{Greenberg}},
  \bibinfo{author}{\bibfnamefont{A.~G.} \bibnamefont{Gavriliuk}},
  \bibnamefont{et~al.}, \emph{\bibinfo{title}{Anomalous high-temperature
  superconductivity in yh$_6$}} (\bibinfo{year}{2020}), \eprint{1908.01534}.

\bibitem[{\citenamefont{Snider et~al.}(2020{\natexlab{a}})\citenamefont{Snider,
  Dasenbrock-Gammon, McBride, Wang, Meyers, Lawler, Zurek, Salamat, and
  Dias}}]{snider2020superconductivity}
\bibinfo{author}{\bibfnamefont{E.}~\bibnamefont{Snider}},
  \bibinfo{author}{\bibfnamefont{N.}~\bibnamefont{Dasenbrock-Gammon}},
  \bibinfo{author}{\bibfnamefont{R.}~\bibnamefont{McBride}},
  \bibinfo{author}{\bibfnamefont{X.}~\bibnamefont{Wang}},
  \bibinfo{author}{\bibfnamefont{N.}~\bibnamefont{Meyers}},
  \bibinfo{author}{\bibfnamefont{K.~V.} \bibnamefont{Lawler}},
  \bibinfo{author}{\bibfnamefont{E.}~\bibnamefont{Zurek}},
  \bibinfo{author}{\bibfnamefont{A.}~\bibnamefont{Salamat}}, \bibnamefont{and}
  \bibinfo{author}{\bibfnamefont{R.}~\bibnamefont{Dias}},
  \emph{\bibinfo{title}{Superconductivity to 262 kelvin via catalyzed
  hydrogenation of yttrium at high pressures}}
  (\bibinfo{year}{2020}{\natexlab{a}}), \eprint{2012.13627}.

\bibitem[{\citenamefont{Kong et~al.}(2019)\citenamefont{Kong, Minkov,
  Kuzovnikov, Besedin, Drozdov, Mozaffari, Balicas, Balakirev, Prakapenka,
  Greenberg et~al.}}]{kong2019superconductivity}
\bibinfo{author}{\bibfnamefont{P.~P.} \bibnamefont{Kong}},
  \bibinfo{author}{\bibfnamefont{V.~S.} \bibnamefont{Minkov}},
  \bibinfo{author}{\bibfnamefont{M.~A.} \bibnamefont{Kuzovnikov}},
  \bibinfo{author}{\bibfnamefont{S.~P.} \bibnamefont{Besedin}},
  \bibinfo{author}{\bibfnamefont{A.~P.} \bibnamefont{Drozdov}},
  \bibinfo{author}{\bibfnamefont{S.}~\bibnamefont{Mozaffari}},
  \bibinfo{author}{\bibfnamefont{L.}~\bibnamefont{Balicas}},
  \bibinfo{author}{\bibfnamefont{F.~F.} \bibnamefont{Balakirev}},
  \bibinfo{author}{\bibfnamefont{V.~B.} \bibnamefont{Prakapenka}},
  \bibinfo{author}{\bibfnamefont{E.}~\bibnamefont{Greenberg}},
  \bibnamefont{et~al.}, \emph{\bibinfo{title}{Superconductivity up to 243 k in
  yttrium hydrides under high pressure}} (\bibinfo{year}{2019}),
  \eprint{1909.10482}.

\bibitem[{\citenamefont{Snider et~al.}(2020{\natexlab{b}})\citenamefont{Snider,
  Dasenbrock-Gammon, McBride, Debessai, Vindana, Vencatasamy, Lawler, Salamat,
  and Dias}}]{Snider2020}
\bibinfo{author}{\bibfnamefont{E.}~\bibnamefont{Snider}},
  \bibinfo{author}{\bibfnamefont{N.}~\bibnamefont{Dasenbrock-Gammon}},
  \bibinfo{author}{\bibfnamefont{R.}~\bibnamefont{McBride}},
  \bibinfo{author}{\bibfnamefont{M.}~\bibnamefont{Debessai}},
  \bibinfo{author}{\bibfnamefont{H.}~\bibnamefont{Vindana}},
  \bibinfo{author}{\bibfnamefont{K.}~\bibnamefont{Vencatasamy}},
  \bibinfo{author}{\bibfnamefont{K.~V.} \bibnamefont{Lawler}},
  \bibinfo{author}{\bibfnamefont{A.}~\bibnamefont{Salamat}}, \bibnamefont{and}
  \bibinfo{author}{\bibfnamefont{R.~P.} \bibnamefont{Dias}},
  \bibinfo{journal}{Nature} \textbf{\bibinfo{volume}{586}},
  \bibinfo{pages}{373} (\bibinfo{year}{2020}{\natexlab{b}}), ISSN
  \bibinfo{issn}{1476-4687},
  \urlprefix\url{https://doi.org/10.1038/s41586-020-2801-z}.

\bibitem[{\citenamefont{Goncharenko et~al.}(2008)\citenamefont{Goncharenko,
  Eremets, Hanfland, Tse, Amboage, Yao, and Trojan}}]{19}
\bibinfo{author}{\bibfnamefont{I.}~\bibnamefont{Goncharenko}},
  \bibinfo{author}{\bibfnamefont{M.~I.} \bibnamefont{Eremets}},
  \bibinfo{author}{\bibfnamefont{M.}~\bibnamefont{Hanfland}},
  \bibinfo{author}{\bibfnamefont{J.~S.} \bibnamefont{Tse}},
  \bibinfo{author}{\bibfnamefont{M.}~\bibnamefont{Amboage}},
  \bibinfo{author}{\bibfnamefont{Y.}~\bibnamefont{Yao}}, \bibnamefont{and}
  \bibinfo{author}{\bibfnamefont{I.~A.} \bibnamefont{Trojan}},
  \bibinfo{journal}{Phys. Rev. Lett.} \textbf{\bibinfo{volume}{100}},
  \bibinfo{pages}{045504} (\bibinfo{year}{2008}),
  \urlprefix\url{https://link.aps.org/doi/10.1103/PhysRevLett.100.045504}.

\bibitem[{\citenamefont{Errea et~al.}(2014)\citenamefont{Errea, Calandra, and
  Mauri}}]{22}
\bibinfo{author}{\bibfnamefont{I.}~\bibnamefont{Errea}},
  \bibinfo{author}{\bibfnamefont{M.}~\bibnamefont{Calandra}}, \bibnamefont{and}
  \bibinfo{author}{\bibfnamefont{F.}~\bibnamefont{Mauri}},
  \bibinfo{journal}{Phys. Rev. B} \textbf{\bibinfo{volume}{89}},
  \bibinfo{pages}{064302} (\bibinfo{year}{2014}),
  \urlprefix\url{https://link.aps.org/doi/10.1103/PhysRevB.89.064302}.

\bibitem[{\citenamefont{Errea et~al.}(2015)\citenamefont{Errea, Calandra,
  Pickard, Nelson, Needs, Li, Liu, Zhang, Ma, and Mauri}}]{23}
\bibinfo{author}{\bibfnamefont{I.}~\bibnamefont{Errea}},
  \bibinfo{author}{\bibfnamefont{M.}~\bibnamefont{Calandra}},
  \bibinfo{author}{\bibfnamefont{C.~J.} \bibnamefont{Pickard}},
  \bibinfo{author}{\bibfnamefont{J.}~\bibnamefont{Nelson}},
  \bibinfo{author}{\bibfnamefont{R.~J.} \bibnamefont{Needs}},
  \bibinfo{author}{\bibfnamefont{Y.}~\bibnamefont{Li}},
  \bibinfo{author}{\bibfnamefont{H.}~\bibnamefont{Liu}},
  \bibinfo{author}{\bibfnamefont{Y.}~\bibnamefont{Zhang}},
  \bibinfo{author}{\bibfnamefont{Y.}~\bibnamefont{Ma}}, \bibnamefont{and}
  \bibinfo{author}{\bibfnamefont{F.}~\bibnamefont{Mauri}},
  \bibinfo{journal}{Phys. Rev. Lett.} \textbf{\bibinfo{volume}{114}},
  \bibinfo{pages}{157004} (\bibinfo{year}{2015}),
  \urlprefix\url{https://link.aps.org/doi/10.1103/PhysRevLett.114.157004}.

\bibitem[{\citenamefont{Errea et~al.}(2016)\citenamefont{Errea, Calandra,
  Pickard, Nelson, Needs, Li, Liu, Zhang, Ma, and Mauri}}]{24}
\bibinfo{author}{\bibfnamefont{I.}~\bibnamefont{Errea}},
  \bibinfo{author}{\bibfnamefont{M.}~\bibnamefont{Calandra}},
  \bibinfo{author}{\bibfnamefont{C.~J.} \bibnamefont{Pickard}},
  \bibinfo{author}{\bibfnamefont{J.~R.} \bibnamefont{Nelson}},
  \bibinfo{author}{\bibfnamefont{R.~J.} \bibnamefont{Needs}},
  \bibinfo{author}{\bibfnamefont{Y.}~\bibnamefont{Li}},
  \bibinfo{author}{\bibfnamefont{H.}~\bibnamefont{Liu}},
  \bibinfo{author}{\bibfnamefont{Y.}~\bibnamefont{Zhang}},
  \bibinfo{author}{\bibfnamefont{Y.}~\bibnamefont{Ma}}, \bibnamefont{and}
  \bibinfo{author}{\bibfnamefont{F.}~\bibnamefont{Mauri}},
  \bibinfo{journal}{Nature} \textbf{\bibinfo{volume}{532}}, \bibinfo{pages}{81}
  (\bibinfo{year}{2016}), ISSN \bibinfo{issn}{1476-4687},
  \urlprefix\url{https://doi.org/10.1038/nature17175}.

\bibitem[{\citenamefont{Errea et~al.}(2020)\citenamefont{Errea, Belli,
  Monacelli, Sanna, Koretsune, Tadano, Bianco, Calandra, Arita, Mauri
  et~al.}}]{25}
\bibinfo{author}{\bibfnamefont{I.}~\bibnamefont{Errea}},
  \bibinfo{author}{\bibfnamefont{F.}~\bibnamefont{Belli}},
  \bibinfo{author}{\bibfnamefont{L.}~\bibnamefont{Monacelli}},
  \bibinfo{author}{\bibfnamefont{A.}~\bibnamefont{Sanna}},
  \bibinfo{author}{\bibfnamefont{T.}~\bibnamefont{Koretsune}},
  \bibinfo{author}{\bibfnamefont{T.}~\bibnamefont{Tadano}},
  \bibinfo{author}{\bibfnamefont{R.}~\bibnamefont{Bianco}},
  \bibinfo{author}{\bibfnamefont{M.}~\bibnamefont{Calandra}},
  \bibinfo{author}{\bibfnamefont{R.}~\bibnamefont{Arita}},
  \bibinfo{author}{\bibfnamefont{F.}~\bibnamefont{Mauri}},
  \bibnamefont{et~al.}, \bibinfo{journal}{Nature}
  \textbf{\bibinfo{volume}{578}}, \bibinfo{pages}{66} (\bibinfo{year}{2020}),
  ISSN \bibinfo{issn}{1476-4687},
  \urlprefix\url{https://doi.org/10.1038/s41586-020-1955-z}.

\bibitem[{\citenamefont{Errea et~al.}(2013)\citenamefont{Errea, Calandra, and
  Mauri}}]{26}
\bibinfo{author}{\bibfnamefont{I.}~\bibnamefont{Errea}},
  \bibinfo{author}{\bibfnamefont{M.}~\bibnamefont{Calandra}}, \bibnamefont{and}
  \bibinfo{author}{\bibfnamefont{F.}~\bibnamefont{Mauri}},
  \bibinfo{journal}{Phys. Rev. Lett.} \textbf{\bibinfo{volume}{111}},
  \bibinfo{pages}{177002} (\bibinfo{year}{2013}),
  \urlprefix\url{https://link.aps.org/doi/10.1103/PhysRevLett.111.177002}.

\bibitem[{\citenamefont{Pickard and Needs}(2007)}]{27}
\bibinfo{author}{\bibfnamefont{C.~J.} \bibnamefont{Pickard}} \bibnamefont{and}
  \bibinfo{author}{\bibfnamefont{R.~J.} \bibnamefont{Needs}},
  \bibinfo{journal}{Phys. Rev. B} \textbf{\bibinfo{volume}{76}},
  \bibinfo{pages}{144114} (\bibinfo{year}{2007}),
  \urlprefix\url{https://link.aps.org/doi/10.1103/PhysRevB.76.144114}.

\bibitem[{\citenamefont{Rousseau and Bergara}(2010)}]{28}
\bibinfo{author}{\bibfnamefont{B.}~\bibnamefont{Rousseau}} \bibnamefont{and}
  \bibinfo{author}{\bibfnamefont{A.}~\bibnamefont{Bergara}},
  \bibinfo{journal}{Phys. Rev. B} \textbf{\bibinfo{volume}{82}},
  \bibinfo{pages}{104504} (\bibinfo{year}{2010}),
  \urlprefix\url{https://link.aps.org/doi/10.1103/PhysRevB.82.104504}.

\bibitem[{\citenamefont{Errea}(2016)}]{Errea2016}
\bibinfo{author}{\bibfnamefont{I.}~\bibnamefont{Errea}}, \bibinfo{journal}{The
  European Physical Journal B} \textbf{\bibinfo{volume}{89}},
  \bibinfo{pages}{237} (\bibinfo{year}{2016}), ISSN \bibinfo{issn}{1434-6036},
  \urlprefix\url{https://doi.org/10.1140/epjb/e2016-70078-6}.

\bibitem[{\citenamefont{Bianco et~al.}(2017)\citenamefont{Bianco, Errea,
  Paulatto, Calandra, and Mauri}}]{29}
\bibinfo{author}{\bibfnamefont{R.}~\bibnamefont{Bianco}},
  \bibinfo{author}{\bibfnamefont{I.}~\bibnamefont{Errea}},
  \bibinfo{author}{\bibfnamefont{L.}~\bibnamefont{Paulatto}},
  \bibinfo{author}{\bibfnamefont{M.}~\bibnamefont{Calandra}}, \bibnamefont{and}
  \bibinfo{author}{\bibfnamefont{F.}~\bibnamefont{Mauri}},
  \bibinfo{journal}{Phys. Rev. B} \textbf{\bibinfo{volume}{96}},
  \bibinfo{pages}{014111} (\bibinfo{year}{2017}),
  \urlprefix\url{https://link.aps.org/doi/10.1103/PhysRevB.96.014111}.

\bibitem[{\citenamefont{Monacelli et~al.}(2018)\citenamefont{Monacelli, Errea,
  Calandra, and Mauri}}]{30}
\bibinfo{author}{\bibfnamefont{L.}~\bibnamefont{Monacelli}},
  \bibinfo{author}{\bibfnamefont{I.}~\bibnamefont{Errea}},
  \bibinfo{author}{\bibfnamefont{M.}~\bibnamefont{Calandra}}, \bibnamefont{and}
  \bibinfo{author}{\bibfnamefont{F.}~\bibnamefont{Mauri}},
  \bibinfo{journal}{Phys. Rev. B} \textbf{\bibinfo{volume}{98}},
  \bibinfo{pages}{024106} (\bibinfo{year}{2018}),
  \urlprefix\url{https://link.aps.org/doi/10.1103/PhysRevB.98.024106}.

\bibitem[{\citenamefont{Monacelli and Mauri}(2020)}]{monacelli2020time}
\bibinfo{author}{\bibfnamefont{L.}~\bibnamefont{Monacelli}} \bibnamefont{and}
  \bibinfo{author}{\bibfnamefont{F.}~\bibnamefont{Mauri}},
  \bibinfo{journal}{arXiv preprint arXiv:2011.14986}  (\bibinfo{year}{2020}).

\bibitem[{\citenamefont{Lihm and Park}(2020)}]{lihm2020gaussian}
\bibinfo{author}{\bibfnamefont{J.-M.} \bibnamefont{Lihm}} \bibnamefont{and}
  \bibinfo{author}{\bibfnamefont{C.-H.} \bibnamefont{Park}},
  \bibinfo{journal}{arXiv preprint arXiv:2010.15725}  (\bibinfo{year}{2020}).

\bibitem[{\citenamefont{Bianco et~al.}(2018)\citenamefont{Bianco, Errea,
  Calandra, and Mauri}}]{PhysRevB.97.214101}
\bibinfo{author}{\bibfnamefont{R.}~\bibnamefont{Bianco}},
  \bibinfo{author}{\bibfnamefont{I.}~\bibnamefont{Errea}},
  \bibinfo{author}{\bibfnamefont{M.}~\bibnamefont{Calandra}}, \bibnamefont{and}
  \bibinfo{author}{\bibfnamefont{F.}~\bibnamefont{Mauri}},
  \bibinfo{journal}{Phys. Rev. B} \textbf{\bibinfo{volume}{97}},
  \bibinfo{pages}{214101} (\bibinfo{year}{2018}),
  \urlprefix\url{https://link.aps.org/doi/10.1103/PhysRevB.97.214101}.

\bibitem[{\citenamefont{Allen and Dynes}(1975)}]{31}
\bibinfo{author}{\bibfnamefont{P.~B.} \bibnamefont{Allen}} \bibnamefont{and}
  \bibinfo{author}{\bibfnamefont{R.~C.} \bibnamefont{Dynes}},
  \bibinfo{journal}{Phys. Rev. B} \textbf{\bibinfo{volume}{12}},
  \bibinfo{pages}{905} (\bibinfo{year}{1975}),
  \urlprefix\url{https://link.aps.org/doi/10.1103/PhysRevB.12.905}.

\bibitem[{\citenamefont{Morel and Anderson}(1962)}]{32}
\bibinfo{author}{\bibfnamefont{P.}~\bibnamefont{Morel}} \bibnamefont{and}
  \bibinfo{author}{\bibfnamefont{P.~W.} \bibnamefont{Anderson}},
  \bibinfo{journal}{Phys. Rev.} \textbf{\bibinfo{volume}{125}},
  \bibinfo{pages}{1263} (\bibinfo{year}{1962}),
  \urlprefix\url{https://link.aps.org/doi/10.1103/PhysRev.125.1263}.

\bibitem[{\citenamefont{Giannozzi et~al.}(2009)\citenamefont{Giannozzi, Baroni,
  Bonini, Calandra, Car, Cavazzoni, Ceresoli, Chiarotti, Cococcioni, Dabo
  et~al.}}]{34}
\bibinfo{author}{\bibfnamefont{P.}~\bibnamefont{Giannozzi}},
  \bibinfo{author}{\bibfnamefont{S.}~\bibnamefont{Baroni}},
  \bibinfo{author}{\bibfnamefont{N.}~\bibnamefont{Bonini}},
  \bibinfo{author}{\bibfnamefont{M.}~\bibnamefont{Calandra}},
  \bibinfo{author}{\bibfnamefont{R.}~\bibnamefont{Car}},
  \bibinfo{author}{\bibfnamefont{C.}~\bibnamefont{Cavazzoni}},
  \bibinfo{author}{\bibfnamefont{D.}~\bibnamefont{Ceresoli}},
  \bibinfo{author}{\bibfnamefont{G.~L.} \bibnamefont{Chiarotti}},
  \bibinfo{author}{\bibfnamefont{M.}~\bibnamefont{Cococcioni}},
  \bibinfo{author}{\bibfnamefont{I.}~\bibnamefont{Dabo}}, \bibnamefont{et~al.},
  \bibinfo{journal}{Journal of Physics: Condensed Matter}
  \textbf{\bibinfo{volume}{21}}, \bibinfo{pages}{395502}
  (\bibinfo{year}{2009}),
  \urlprefix\url{https://doi.org/10.1088%2F0953-8984%2F21%2F39%2F395502}.

\bibitem[{\citenamefont{Giannozzi et~al.}(2017)\citenamefont{Giannozzi,
  Andreussi, Brumme, Bunau, Nardelli, Calandra, Car, Cavazzoni, Ceresoli,
  Cococcioni et~al.}}]{Giannozzi_2017}
\bibinfo{author}{\bibfnamefont{P.}~\bibnamefont{Giannozzi}},
  \bibinfo{author}{\bibfnamefont{O.}~\bibnamefont{Andreussi}},
  \bibinfo{author}{\bibfnamefont{T.}~\bibnamefont{Brumme}},
  \bibinfo{author}{\bibfnamefont{O.}~\bibnamefont{Bunau}},
  \bibinfo{author}{\bibfnamefont{M.~B.} \bibnamefont{Nardelli}},
  \bibinfo{author}{\bibfnamefont{M.}~\bibnamefont{Calandra}},
  \bibinfo{author}{\bibfnamefont{R.}~\bibnamefont{Car}},
  \bibinfo{author}{\bibfnamefont{C.}~\bibnamefont{Cavazzoni}},
  \bibinfo{author}{\bibfnamefont{D.}~\bibnamefont{Ceresoli}},
  \bibinfo{author}{\bibfnamefont{M.}~\bibnamefont{Cococcioni}},
  \bibnamefont{et~al.}, \bibinfo{journal}{Journal of Physics: Condensed Matter}
  \textbf{\bibinfo{volume}{29}}, \bibinfo{pages}{465901}
  (\bibinfo{year}{2017}),
  \urlprefix\url{https://doi.org/10.1088/1361-648x/aa8f79}.

\bibitem[{\citenamefont{Vanderbilt}(1990)}]{35}
\bibinfo{author}{\bibfnamefont{D.}~\bibnamefont{Vanderbilt}},
  \bibinfo{journal}{Phys. Rev. B} \textbf{\bibinfo{volume}{41}},
  \bibinfo{pages}{7892} (\bibinfo{year}{1990}),
  \urlprefix\url{https://link.aps.org/doi/10.1103/PhysRevB.41.7892}.

\bibitem[{\citenamefont{Perdew et~al.}(1996)\citenamefont{Perdew, Burke, and
  Ernzerhof}}]{36}
\bibinfo{author}{\bibfnamefont{J.~P.} \bibnamefont{Perdew}},
  \bibinfo{author}{\bibfnamefont{K.}~\bibnamefont{Burke}}, \bibnamefont{and}
  \bibinfo{author}{\bibfnamefont{M.}~\bibnamefont{Ernzerhof}},
  \bibinfo{journal}{Phys. Rev. Lett.} \textbf{\bibinfo{volume}{77}},
  \bibinfo{pages}{3865} (\bibinfo{year}{1996}),
  \urlprefix\url{https://link.aps.org/doi/10.1103/PhysRevLett.77.3865}.

\bibitem[{\citenamefont{Baroni et~al.}(2001)\citenamefont{Baroni, de~Gironcoli,
  Dal~Corso, and Giannozzi}}]{DFPT_S.Baroni}
\bibinfo{author}{\bibfnamefont{S.}~\bibnamefont{Baroni}},
  \bibinfo{author}{\bibfnamefont{S.}~\bibnamefont{de~Gironcoli}},
  \bibinfo{author}{\bibfnamefont{A.}~\bibnamefont{Dal~Corso}},
  \bibnamefont{and}
  \bibinfo{author}{\bibfnamefont{P.}~\bibnamefont{Giannozzi}},
  \bibinfo{journal}{Rev. Mod. Phys.} \textbf{\bibinfo{volume}{73}},
  \bibinfo{pages}{515} (\bibinfo{year}{2001}),
  \urlprefix\url{http://link.aps.org/doi/10.1103/RevModPhys.73.515}.

\end{thebibliography}

\end{document}